\RequirePackage{snapshot}
\documentclass[letterpaper,twocolumn,10pt]{article}
\usepackage{usenix2019}
\usepackage{authblk}

\newcommand*{\affaddr}[1]{#1} 
\newcommand*{\affmark}[1][*]{\textsuperscript{#1}}

\usepackage{wrapfig}
\usepackage{comment}
\usepackage{colortbl}
\usepackage{boldline}

\usepackage{amssymb}
\usepackage{pifont}
\usepackage{makecell}

\usepackage{hyperref}

\usepackage{color,soul}
\usepackage{graphics}
\usepackage{hyperref}
\usepackage{lipsum}
\usepackage{tikz}
\usepackage[inline]{enumitem} 

\usepackage{listings} 
\usepackage{booktabs}	
\usepackage{lipsum}

\usepackage{capt-of}	

\usepackage{todonotes}  
\usepackage{subcaption} 

\usepackage{mathrsfs}	
\usepackage{amsfonts}
\usepackage{inconsolata} 
\usepackage{pdfcomment} 

\usepackage[export]{adjustbox} 

\usepackage{mathtools} 

\renewcommand{\paragraph}[1]{\vskip 3pt\noindent\textbf{#1 }}	 

\newenvironment{smitemize}%
  {\begin{list}{$\bullet$}%
     {\setlength{\parsep}{0pt}%
      \setlength{\topsep}{0pt}%
      \setlength{\itemsep}{2pt}}}%
  {\end{list}}
\newcommand\note[1]{\sethlcolor{yellow} \hl{#1}} 
						 
\newcommand\noted[1]{} 

\newcommand\hp[1]{{\color{blue}{#1}}}
\newcommand\hpsout[1]{{\color{red}\sout{hp: #1}}}

\newcommand\sect[1]{Section~\ref{sec:#1}}

\newenvironment{myitemize}%
  {\begin{itemize}
	[leftmargin=0cm,
		itemindent=.3cm,
		labelwidth=\itemindent,
		labelsep=0pt,
		parsep=3pt,
		topsep=2pt,
		itemsep=1pt,
		align=left]
  }%
  {\end{itemize}}    

\newenvironment{myenumerate}%
  {\begin{enumerate}
	[leftmargin=0cm,itemindent=.5cm,labelwidth=\itemindent,
		labelsep=0pt,
		parsep=1pt,
		topsep=1pt,
		itemsep=3pt,
		align=left]
  }%
  {\end{enumerate}}    

\newenvironment{enumerateinline}%
  {\begin{enumerate*}
	[label=\roman*)]
  }%
  {\end{enumerate*}}      
  
\newcommand\textbfit[1]{\textit{\textbf{#1}}}

\newcommand{\code}[1]{\texttt{\small{#1}}}	

\newcommand{\se}{secure world}

\newcommand{\nw}{normal world}
\newcommand{\sw}{secure world}

\newcommand{\ubuf}{uArray}

\newcommand{\bufgroup}{uGroup}

\newcommand{\sysfull}{StreamBox-TZ}
\newcommand{\sys}{SBT}
\newcommand{\engine}{SBT}

\newcommand{\tas}{primitive}
\newcommand{\task}{trusted primitive}
\newcommand{\Task}{Trusted primitive}
\newcommand{\TASK}{Trusted Primitive}

\newcommand{\TZ}{TrustZone}

\newcommand{\tee}{trusted execution environment}

\newcommand{\spe}{stream analytics engine}
\newcommand{\Spe}{Stream analytics engine}

\newcommand{\filter}{Filter}
\newcommand{\winsum}{WinSum}
\newcommand{\join}{Join}
\newcommand{\unique}{Distinct}
\newcommand{\freq}{TopK}
\newcommand{\grid}{Power}

\newcommand{\sender}{\textit{Generator}}

\newcommand{\unsecure}{\textit{Insecure}}
\newcommand{\sysindenc}{\textit{SBT IOviaOS}}
\newcommand{\sysdirclr}{\textit{SBT ClearIngress}}
\newcommand{\sysdirenc}{\textit{\sys{}}}

\makeatletter
\def\@copyrightspace{\relax}
\makeatother

\begin{document}
\title{\sysfull{}: Secure Stream Analytics at the Edge with TrustZone}



\author{
	\rm Heejin Park\affmark[1], 
	\rm Shuang Zhai\affmark[1], 
	\rm Long Lu\affmark[2], and 
	\rm Felix Xiaozhu Lin\affmark[1]\\
	\affaddr{\affmark[1]Purdue ECE~~~~~~~~~\affmark[2]Northeastern University}
}


\date{}           
\maketitle
\thispagestyle{empty}
\pagestyle{empty}


\subsection*{Abstract}

While it is compelling to process large streams of IoT data on the cloud edge,
doing so exposes the data to a sophisticated, vulnerable software stack on the edge and hence security threats. 
To this end, we advocate isolating the data and its computations in a trusted execution environment (TEE) on the edge, shielding them from the remaining edge software stack which we deem untrusted. 

This approach faces two major challenges:
(1) executing high-throughput, low-delay stream analytics in a single TEE, which is constrained by 
a low trusted computing base (TCB) and limited physical memory;
(2) verifying execution of stream analytics as the execution involves untrusted software components on the edge. 
In response, we present \sysfull{} (\sys{}), a stream analytics engine for an edge platform that offers strong data security, verifiable results, and good performance.
\sys{} contributes a data plane designed and optimized for a TEE based on ARM TrustZone. 
It supports continuous remote attestation for analytics correctness and result freshness while incurring low overhead. 
\sys{} only adds 42.5~KB executable to the TCB (16\% of the entire TCB).
On an octa core ARMv8 platform, it delivers the state-of-the-art performance by processing input events up to 140 MB/sec (12M events/sec) with sub-second delay.
The overhead incurred by \sys{}'s security mechanism is less than 25\%.

\newcommand{\isolation}{Exploiting unique advantages of ARM TrustZone}
\newcommand{\noconcurrency}{Fast computations in a lean TCB}

\newcommand{\seqmm}{Specializing memory abstractions \& management}

\newcommand{\hints}{Attesting analytics results by cloud server}

\section{Introduction}
\label{sec:intro}



Many key applications of Internet of Things (IoT) process a large influx of sensor\footnote{Recognizing that IoT data sources range from small sensors to large equipment, we refer to them all as \textit{sensors} for brevity.} data, i.e. telemetry. 
Smart grid aggregates power telemetry to detect supply/demand imbalance and power disturbances~\cite{smart-grid}, where a power sensor is reported to produce up to 140 million samples per day~\cite{distil, btrdb};
oil producers monitor pump pressure, tank status, and fluid temperatures to determine if wells work at ideal operating points~\cite{intel-oil, hortonworks}, where an oil rig is reported to produce 1--2 TB of data per day~\cite{cisco-oil};
manufacturers continuously monitor vibration and ultrasonic energy of industrial equipment for detecting equipment anomaly and predictive maintenance~\cite{oracle-predictive, worx}, where
a monitored machine is reported to generate PBs of data in a few days~\cite{dell-manufacturing}.


The large telemetry data streams must be processed in time. 
The high cost and long delay in transmitting data necessitate edge processing~\cite{shi16iotj,satya-edge-iot}: 
sensors send the data to nearby gateways dubbed ``cloud edge'';
the edge runs a pipeline of continuous computations to cleanse and summarize the  telemetry data and reports the results to cloud servers for deeper analysis. 
Edge hardware is often optimized for cost and efficiency. 
According to a 2018 survey~\cite{iot-survey}, modern ARM machines are typical choices for edge platforms. 
Such a platform often has 2--8 CPU cores and several GB DRAM. 

\begin{figure}[t!]
\centering
\includegraphics[width=0.45\textwidth{}]{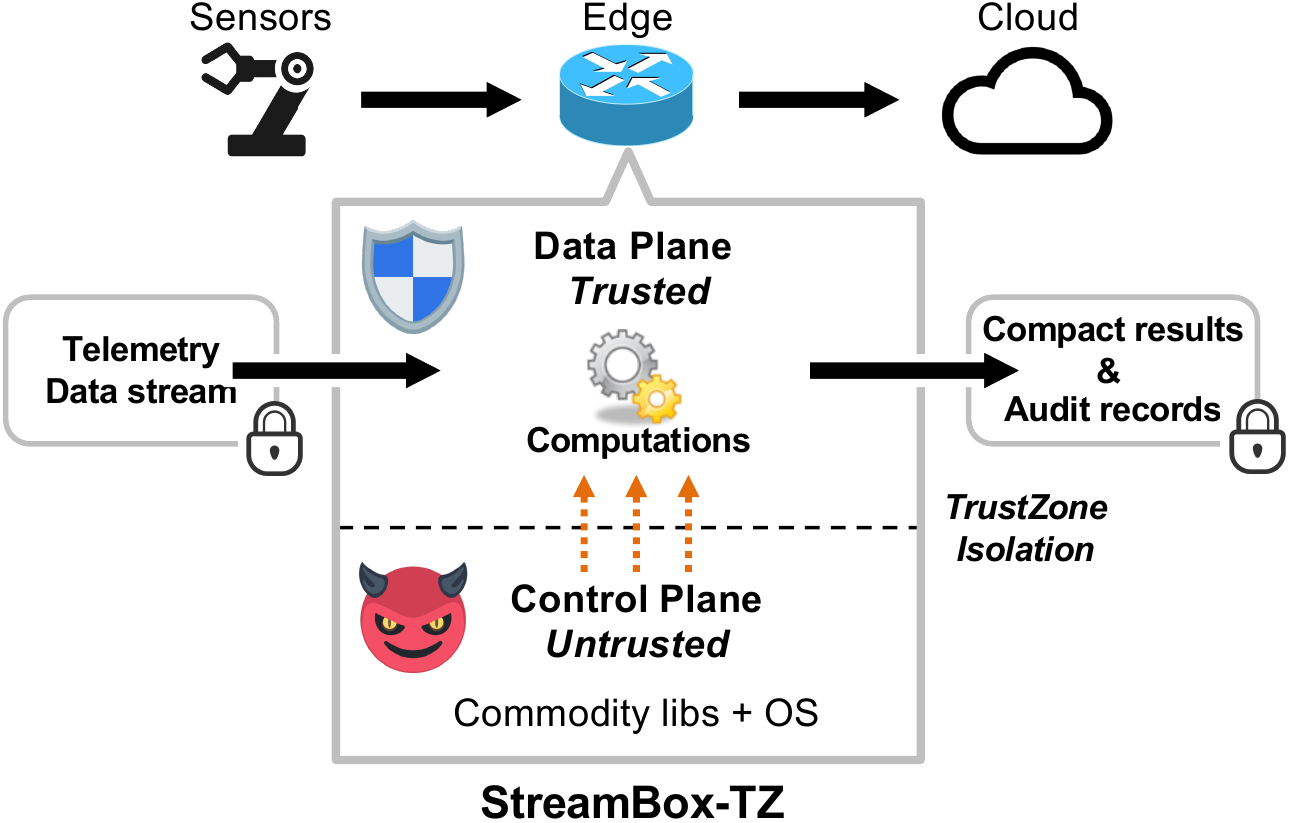} 
\vspace{-9pt}		
\caption{An overview of \sysfull{}}
\label{fig:overview}
\vspace{-18pt}		
\end{figure}

Unfortunately, edge processing exposes IoT data to high security threats.
i)
Deployed in the wild, the edge suffers from common IoT weaknesses,
 including lack of professional supervision~\cite{mckinsey-iot,Wu:2017:ECE:3132465.3132475}, weak configurations~\cite{symantec-iot-security,windriver-iot-security}, and long delays in receiving security updates~\cite{farmbeats,mckinsey-iot}. 
ii) 
On the edge, the IoT data flows through a set of sophisticated components that expose a wide attack surface. 
These components include a commodity OS (e.g. Linux or Windows), a variety of user libraries, and a runtime framework called \textit{\spe{}}~\cite{msft-azure-edge,cisco-edge-fabric,dell-statistica}. 
They reuse much code developed for servers and workstations.
Their exploitable misconfigurations \cite{misconfig-sosp11} and vulnerabilities~\cite{anatomy-java-exploits,chen2011linux,tan2014bug} are not uncommon.
iii) With data aggregated from multiple sources, the edge is a high-value target to adversaries.
For these reasons, edge is even more vulnerable than sensors, which run much simpler software with narrower attack surfaces. 
Once attackers compromise the edge, they not only access confidential data but also may delete or fabricate data sent to the cloud, threatening the integrity of an entire IoT deployment.

Towards secure stream analytics on an edge platform, 
our goal is to safeguard IoT data confidentiality and integrity, support verifiable results, and ensure high throughput with low output delay.
Following the principle of least privilege~\cite{least-privilege},
we protect the analytics data and computations in a \tee{} (TEE) and limit their interface; 
we leave out the remaining edge software stack which we deem untrusted. 
By doing so, we shrink the trusted computing base (TCB) to only the protected  functionalities, the TEE, and the hardware. We hence significantly enhance data security.

We face three challenges:
i) what functionalities should be protected in TEE and behind what interfaces? 
ii) how to execute stream analytics on a TEE's low TCB and limited physical memory while still delivering high throughput and low delay? 
iii) as both trusted and untrusted edge components participate in stream analytics,
how to verify the outcome?



Existing solutions are inadequate:
pulling entire stream analytics engines to TEE~\cite{scone,graphene-sgx,haven}
 would result in a large TCB with a wide attack surface;
the systems securing distributed operators~\cite{vc3,opaque,securestreams} 
often lack stream semantics or optimizations for efficient execution in a single TEE, which are crucial to the edge; 
only attesting TEE integrity~\cite{sgx-tor} or data lineages~\cite{vc3,opaque,ariadne,data-provenance} is inadequate for verifying stream analytics.
We will show more evidences in the paper. 

Our response is \sysfull{} (\sys{}), a secure engine for analyzing telemetry data streams.
As shown in Figure~\ref{fig:overview}, \sys{} builds on ARM \TZ{}~\cite{arm-trustzone} on an edge platform. 
\sys{} contributes the following notable designs:

\noindent
\textit{(1) Architecting a data plane for protection} \hspace{1mm}
\sys{} provides a data plane exposing narrow, shared-nothing interfaces to untrusted software. 
\sys{}'s data plane encloses i) all the analytics data; 
ii) a new library of low-level stream algorithms called \textit{\task{}s} as the only allowed computations on the data;
iii) key runtime functions, including memory management and cache-coherent parallel execution of \task{}s. 
\sys{} leaves thread scheduling and synchronization out of TEE. 

\noindent
\textit{(2) Optimizing data plane performance within a TEE} \hspace{1mm}  
In contrast to many TEE-oblivious stream engines that operate numerous small objects, hash tables, and generic memory allocators~\cite{sparkstreaming,streambox,trill},
\sys{} embraces unconventional design decisions for its data plane.
i) \sys{} implements \task{}s with array-based algorithms and contributes new optimizations with handwritten ARMv8 vector instructions. 
ii)
To process high-velocity data in TEE,
\sys{} provides a new abstraction called \ubuf{}s, which are contiguous, virtually unbounded buffers for encapsulating all the analytics data;
\sys{} backs \ubuf{}s with on-demand paging in TEE and manages \ubuf{}s with a specialized allocator.
The allocator leverages hints from untrusted software for compacting memory layout.
iii)
\sys{} exploits \TZ{}'s lesser-explored hardware features:
ingesting data straightly through trusted IO without a detour through the untrusted OS;
avoiding relocating streaming data by leveraging the large virtual address space dedicated to a TEE.  

\noindent
\textit{(3) Verifying edge analytics execution} \hspace{1mm}
\sys{} supports cloud verifiers to attest analytics \textit{correctness}, result \textit{freshness}, and the untrusted hints received during execution. 
\sys{} captures coarse-grained dataflows and generates audit records.
A cloud verifier replays the audit records for attestation.
To minimize overhead in the edge-cloud uplink bandwidth, 
\sys{} compresses the records with domain-specific encoding.

Our implementation of \sys{} supports a generic stream model~\cite{beam} with a broad arsenal of stream operators.
The TCB of \sys{} contains as little as 267.5~KB of executable code, of which \sys{} only constitutes 16\%.
%
On an octa core ARMv8 platform, \engine{} processes up to 12M events (144 MB) per second at sub-second output delays.
Its throughput on this platform is an order of magnitude higher than an SGX-based secure stream engine running on a small x86 cluster with richer hardware resources~\cite{securestreams}.
The security mechanisms contributed by \sys{} incur less than 25\% throughput loss with the same output delay;
decrypting ingress data, when needed, incurs 4\%--35\% throughput loss with the same output delay.
While sustaining high throughput, \sys{} uses up to 130 MB of physical memory in most benchmarks.

The key contributions of \sys{} are: 
i) a stream engine architecture with strongly isolated data and a lean TCB;
ii) a data plane built from the ground up with computations and memory management optimized for a single TrustZone-based TEE;
iii) remote attestation for stream analytics on the edge with domain-specific compression of audit records. 
To our knowledge, 
\sys{} is the first system designed and optimized for data-intensive, parallel computations inside ARM TrustZone. 
Beyond stream analytics, the \sys{} architecture should help secure other important analytics on the edge, e.g. machine learning inference.
The \engine{} source can be found at \code{http://xsel.rocks/p/streambox}.

\section{Background \& Motivation}
\label{sec:bkgnd}


\subsection{ARM for Cloud Edge} 
As typical hardware for IoT gateways~\cite{iot-survey},
recent ARM platforms offer competitive performance at low power, suiting edge well. 
Most modern ARM cores are equipped with TrustZone~\cite{arm-trustzone}, a security extension for TEE enforcement. 
TrustZone logically partitions a platform's hardware resources, e.g. DRAM and IO, into a normal (insecure) and a secure world.
CPU cores independently switch between two worlds.
A TEE atop \TZ{} owns dedicated, \textit{trusted IO}, a unique feature that other TEE technologies such as Intel SGX~\cite{intel-sgx} lack.


\noindent \textbf{Trusted IO} is a unique feature of ARM TrustZone, implemented through hardware components including TrustZone Address Space Controller (TZASC) and TrustZone Protection Controller (TZPC).
TZASC allows privilege software to logically partition DRAM between the normal and the secure worlds.
Similarly, TZPC allows to configure IO peripherals accessible to either world.
Any peripheral owned by the secure world is completely enclosed in the secure world. 
We use trusted IO to support the trusted source-edge links on the cloud edge (\S\ref{sec:overview:scope}).




\subsection{Stream Analytics}
\label{sec:motiv:data}
\label{sec:bkgnd:pipeline}


\begin{figure}[t!]
\centering
   \vspace{2pt}
   	\begin{subfigure}[b]{0.4\textwidth}
   \includegraphics[width=1\linewidth]{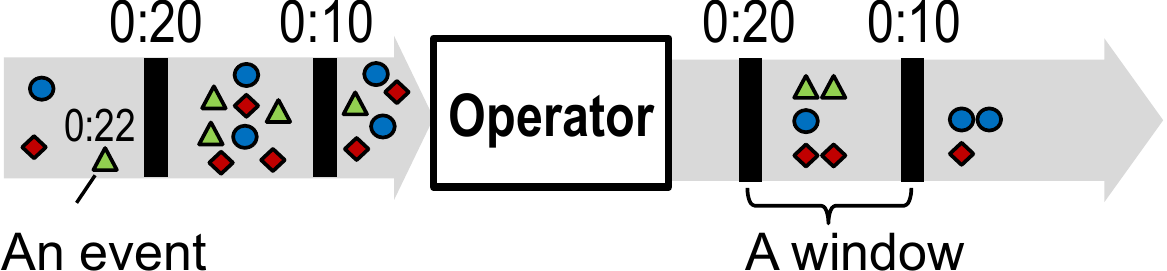}
   \vspace{-5.5mm}
   \caption{A stream of events flowing through an operator.
}
	\vspace{1mm}
   \label{fig:pipeline-stacked:pipeline} 
	\end{subfigure}

	\begin{subfigure}[b]{0.45\textwidth}
	   \includegraphics[width=1\linewidth]{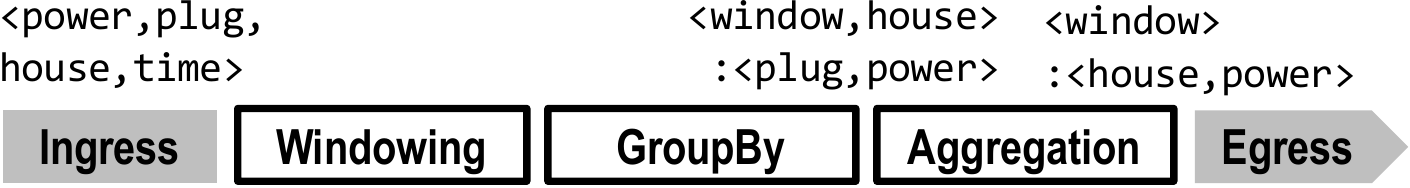}
		\vspace{-5.5mm}
	   \caption{A simple analytics pipeline that predicts power grid loads}
	\label{fig:pipeline-stacked:stream} 
	\end{subfigure}
	
	\begin{subfigure}[b]{0.45\textwidth}
		\input{recycle-code}
		\vspace{-3.5mm}
	   \caption{Simplified pseudo code declaring the above pipeline}	   
  		\vspace{-2mm}
	\label{fig:pipeline-stacked:op} 
	\end{subfigure}

\vspace{-3pt}		
\caption{Example stream data, operators, and a pipeline}
\vspace{-16pt}		

\label{fig:pipeline-stacked}
\end{figure}

\paragraph{Stream Model}
We target stream analytics over sensor data. 
A data stream consists of sensor events that carry timestamps defined by event occurrence, 
as illustrated in Figure~\ref{fig:pipeline-stacked}(a).
Programmers specify a pipeline of continuous computations called \emph{operators}, e.g. \code{Select} and \code{GroupBy}, that are extensively used for telemetry analytics~\cite{sensorbee,smart-power-plugs}.
As data arrives at the edge, a \spe{} ingests the data at the pipeline ingress, pushes the data through the pipeline, and externalizes the results at the pipeline egress.
 
We follow a generic stream model~\cite{naiad,sparkstreaming,dataflowAkidau2015,trill,streamscope}. 
Operators execute on event-time scopes called \textit{windows}.
Data sources emit special events called \textit{watermarks}. 
A watermark guarantees no subsequent events in the stream will have event times earlier than the watermark timestamp. 
A pipeline's \textit{output delay} is defined as the elapsed time starting from the moment the ingress receives the watermark signaling the completion of the current window to the moment the egress externalizes the window results~\cite{streambox}. 
A pipeline may maintain its internal states organized by windows at different operators. 
See prior work~\cite{cql} for a formal stream model.

\paragraph{Analytics example: Power load prediction}
Figure~\ref{fig:pipeline-stacked}(b-c) shows an example derived from an IoT scenario~\cite{smart-power-plugs}:
it predicts future household power loads based on power loads reported by smart power plugs.
The example pipeline ingests a stream of power samples and groups them by 1-second fixed windows and by houses. 
For each house in each window, it aggregates all the loads and predicts the next-window load as an exponentially weighted moving average over the recent windows. 
At the egress, the pipeline emits a stream of per-house load prediction for each window. 

\paragraph{\Spe{}s}
\label{sec:bkgnd:engine}
Stream pipelines are executed by a runtime framework called a \spe{}~\cite{esper,sensorbee,msft-azure-edge,cisco-edge-fabric,dell-statistica,streambox}. 
A \spe{} consists of two types of function:
\textit{data functions} for data move and computations;
\textit{control functions} for 
resource management and computation orchestration, e.g. creating and scheduling tasks.
The boundary between the two is often blurry. 
To amortize overheads, control functions often organize data in \textit{batches} and invoke data functions to operate on the batches. 







\subsection{Security Threats \& Design Objectives}

The edge faces common threats in IoT deployment. 
i) IT expertise is weak. 
Edge platforms are likely managed by field experts~\cite{farmbeats,Wu:2017:ECE:3132465.3132475,mckinsey-iot} rather than IT experts. 
Such lack of professional supervision is known to result in weak configurations~\cite{symantec-iot-security,windriver-iot-security}.
ii) The infrastructure is weak.
Deployed in the field, the edge often sees slow uplinks~\cite{farmbeats,satellites-iot} and hence much delayed software security updates.
For cost saving, edge analytics may need to share OS and hardware with other high-risk, untrusted software such as web browsers~\cite{farmbeats}.

Besides the common threats, existing edge software stacks entrust IoT data with commodity OSes, analytics engines, and language runtimes (e.g. JVM). However, these components are incapable of offering strong security guarantees due to their complexity and wide interfaces. Each of them easily contains more than several hundreds of KSLoC~\cite{relay-ESEC-FSE07}. Exploitable vulnerabilities are constantly discovered~\cite{anatomy-java-exploits, golang-security, CVE-2010-3190, CVE-2008-0171, chen2011linux}, making these components untrusted in recent research~\cite{overshadow, trustvisor, Flicker, inktag}. By exploiting these vulnerabilities, a local adversary as an edge user program may compromise the kernel through the wide user/kernel interfaces~\cite{CVE-2018-8822, CVE-2017-11176} or attack an analytics engine through IPC~\cite{CVE-2009-2493}; a remote adversary, through the edge's network services, may compromise analytics engines~\cite{CVE-2017-12629} or the OS~\cite{CVE-2016-10229}.
A successful adversary may expose IoT data, corrupt the data, or covertly manipulate the data. Taking the application in Figure~\ref{fig:pipeline-stacked}(b) as an example, the adversary gains access to the smart plug readings, which may contain residents' private information, and injects fabricated data.

\paragraph{Objectives}
We aim three objectives for stream analytics over telemetry data on an edge platform:
i) confidentiality and integrity of IoT data, raw or derived;
ii) verifiable correctness and freshness of the analytics results; 
iii) modest security overhead and good performance.

\section{Security Approach Overview}
\label{sec:bkgnd:secmodel}
\label{sec:motiv:objective}


\subsection{Scope}
\label{sec:overview:scope}
\paragraph{IoT scenarios}
We target an edge platform that captures and analyzes telemetry data. 
We recognize the significance of mission-critical IoT with tight control loops, but do not target it.
Our target scenario includes source sensors, edge platforms, and a cloud server which we dub ``cloud consumer''.
All the raw IoT data and analytics results are owned by one party. 
The sensors produce trusted events, e.g. by using secure sensing techniques~\cite{IamSensor,sw_abstractions,YouProve}.
The cloud consumer is trusted; it installs analytics pipelines to the edge and consumes the results uploaded from the edge.
We consider \textit{untrusted} source-edge links (e.g. public networks) which requires data encryption by the source, as well as \textit{trusted} source-edge links (e.g. direct IO bus or on-premise local networks), and will evaluate the corresponding designs (\S\ref{sec:eval}).
We assume untrusted edge-cloud links, which require encryption of the uploaded data. 

\paragraph{In-scope Threats}
We consider malicious adversaries interested in learning IoT data, tampering with edge processing outcome, or obstructing processing progress. 
We assume powerful adversaries: by exploiting weak configurations or bugs in the edge software, they already control the entire OS and all applications on the edge.


\paragraph{Out-of-scope Threats}
We do not protect the confidentiality of stream pipelines, in the interest of including only low-level compute primitives in a lean TCB.
We do not defend the following attacks.
\begin{enumerateinline}

\item
Attacks to non-edge components assumed trusted above, e.g. sensors~\cite{walnut}.

\item
Exploitation of TEE kernel bugs~\cite{vtz, CVE-2015-4421, CVE-2015-4422}. 

\item
Side channel attacks:
by observing hardware usage outside TEE,
adversaries may learn the properties of protected data, e.g. key skew~\cite{armageddon}.
Note that controlled-channel attack~\cite{controlledChannel} cannot be applied to ARM TrustZone 
as it has separate page management within a separate secure OS unlike Intel SGX.

\item
Physical attacks, e.g. sniffing TEE's DRAM access~\cite{tampering,lowCostAttack}. 
Many of these attacks are mitigated by prior work~\cite{sel4,opaque,sentry,TZ-CaSE} orthogonal to \sys{}.


\end{enumerateinline}

Note that TEE code authenticity and integrity are already ensured by the TrustZone hardware, i.e. only code trusted by the device vendor can run in TrustZone and its integrity is protected by TrustZone.

\begin{figure}[t!]
\centering
\vspace{1pt}		
\includegraphics[width=0.35\textwidth{}]{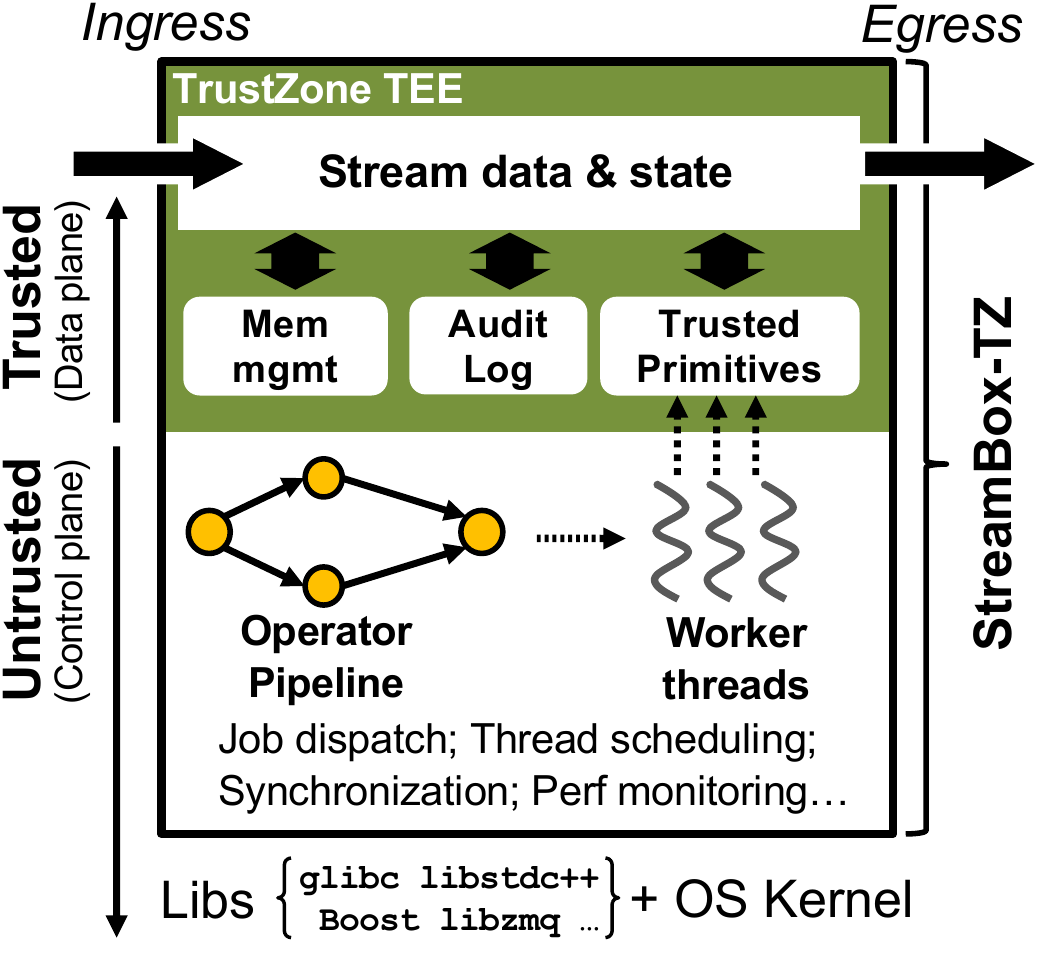} 
\vspace{-11pt}		
\caption{\sysfull{} on an edge platform with ARM \TZ{}. Bold arrows show the protected data path.}
\label{fig:arch}
\vspace{-17pt}		
\end{figure}


%
%
%

\begin{table*}[]
	\centering
    \begin{minipage}[]{0.35\textwidth}
	    \centering
	    \vspace{11pt}
	    \includegraphics[width=1.00\textwidth]{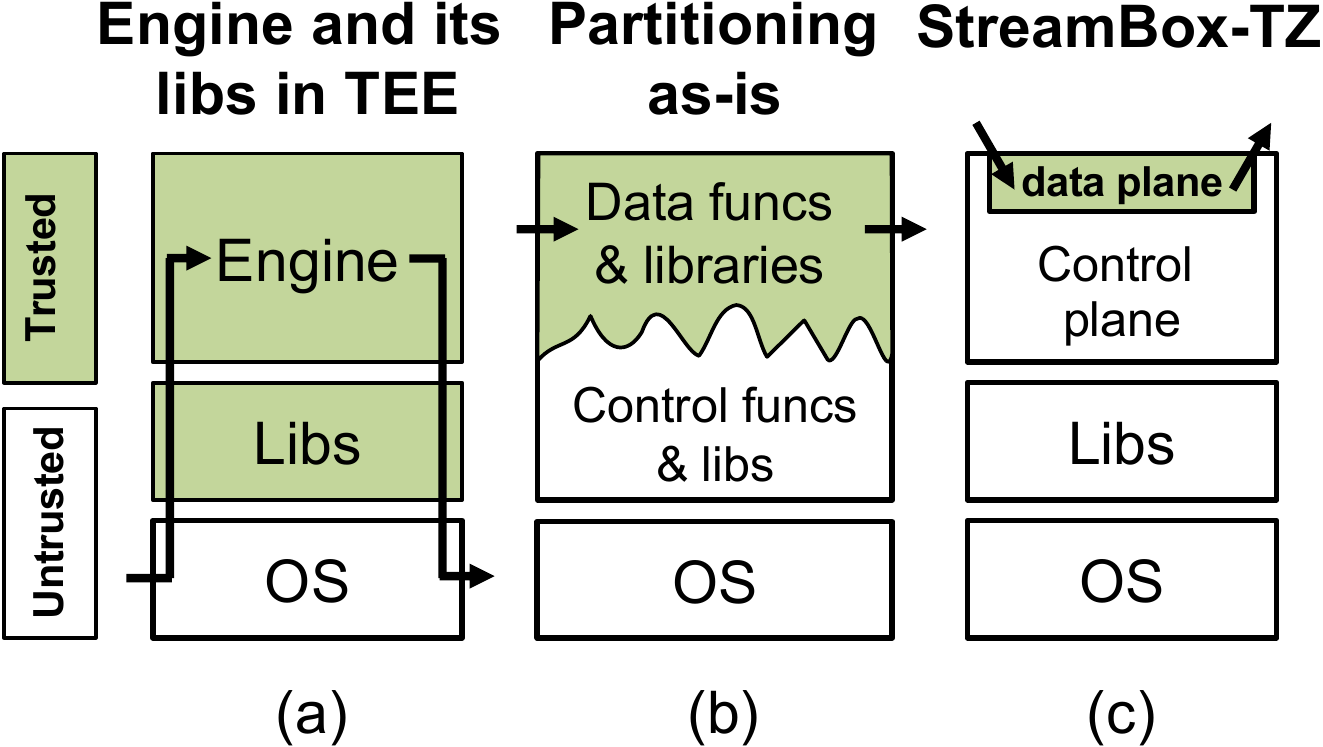}
	    \vspace{-11pt}		
	    \captionof{figure}{Among alternative architectures for secure stream analytics,
	    	\sysfull{} (c) leads to the smallest TCB and the most optimized data plane. Arrows indicate data flows.}
	    \label{fig:div}
	\end{minipage}
	~
    \begin{minipage}[]{0.625\textwidth}
		\vspace{-3pt}
	\fontsize{8.0}{9.0}\selectfont
	\begin{center}
		\setlength\tabcolsep{3pt}	
		\begin{tabular}{
				l	
				l	
				l	
				l	
				l	
				l	
				l	
			}
			\hlineB{2.5}
			\cellcolor{lightgray}\textbf{System} &
			\cellcolor{lightgray}\textbf{TEE} & 
			\cellcolor{lightgray}\textbf{Analytics} &
			\cellcolor{lightgray}\textbf{SG} &
			\cellcolor{lightgray}\textbf{Compute in TEE} &
			\cellcolor{lightgray}\textbf{Memory} &
			\cellcolor{lightgray}\textbf{Attestation}
			\\ 
			\hlineB{1.5}
VC3~\cite{vc3} 				 	
& SGX
& Batch
& \texttt{CIVA-} 
& Mapper/reducer
& Heap
& Data lineage\\

Opaque~\cite{opaque}
& SGX
& Batch
& \texttt{CIVAO}
& Query plans
& unreported
& Data lineage \\

EnclaveDB~\cite{enclaveDB}
& SGX
& Batch
& \texttt{CI-A-}
& Pre-compiled queries
& unreported 
& TEE integrity \\

SafeBricks~\cite{safebricks}
& SGX
& Pkt proc.
& \texttt{CI-A-}
& Net func. operators\textsuperscript{$\ast$}
& unreported 
& TEE integrity \\ 

SecureStream~\cite{securestreams}
& SGX
& Stream
& \texttt{CI---} 
& Lua programs
& unreported 
& TEE integrity\\

\sysfull{}
& TZ
& Stream
& \texttt{CIV--}
& Vectorized primitives\textsuperscript{$\ast$}
& uArray
& Log replay\\

\hlineB{2.5}
		\end{tabular}
		\fontsize{7.3}{7.7}\selectfont
		\textbf{SG}: security guarantees. \\ 
		\texttt{C}: data confidentiality; 
		\texttt{I}: data integrity;  
		\texttt{V}: verifiability; 
		\texttt{A}: analytics confidentiality;
		\texttt{O}: obliviousness \\
		* TEE encloses only low-level computations; otherwise TEE encloses whole analytics.
		\vspace{-3.5mm}
		\caption{Comparison to existing secure processing systems}
		\label{tab:cmp}
	\end{center}

		\vspace{-7mm}
		\centering
		\fontsize{7.6}{8.5}\selectfont
		\begin{center}
			\begin{tabular}{
					c	
					c	
				}
				\hlineB{2.5}
				\cellcolor{lightgray}\textbf{Trusted Primitives} &
				\cellcolor{lightgray}\textbf{Popular Spark Streaming Operators} \\
				\hlineB{1.5}
				\makecell{
					Sort, Merge, Segment, SumCnt, \\
					TopK, Concat, Join, Count, Sum, \\
					Unique, FileterBand, Median, ...
				}
				&
				\makecell{
					GroupByKey, Windowing, AvgPerKey, Distinct, SumByKey, \\
					AggregateByKey, SortByKey, TopKPerKey, CountByKey, \\
					CountByWindow, Filter, MedianByKey, TempJoin, Union, ... 
				} \\
				\hlineB{2.5}
			\end{tabular}
		\end{center}

		\vspace{-6.5mm}
		\caption{Selected \task{}s (23 in total) and operators they constitute. 
		These operators cover most listed in the Spark Streaming documentation~\cite{spark-doc}.
		}
		\label{tab:operators}
		\vspace{-3mm}
	\end{minipage}
	\vspace{-6mm}
\end{table*}
\subsection{Approach and Security Benefits}
As shown in Figure~\ref{fig:arch}, 
\sys{} protects its data functions in a trusted \textit{data plane} in TEE. 
\sys{} runs its untrusted \textit{control plane} in the \nw{}. 
The control plane invokes the data plane through narrow, shared-nothing interfaces.
The engine's TCB thus only consists of the TEE (including the data plane) and the hardware. 

Streaming data always flows in TEE. 
The data plane ingests the data through \TZ{}'s trusted IO.
After ingestion, it returns \textit{opaque references} of the data batches to the control plane. 
In turn, the control plane requests computations on the protected data by invoking the data plane with the opaque references.
The data plane generates opaque references as long, random integers.
It tracks all live opaque references, validates incoming opaque references, and only accepts ones that exist. 
At the pipeline egress, the data plane encrypts, signs, and sends the result to the cloud.


The analytics execution is continuously attested.
\sys{} captures complete and deterministic dataflows of the stream analytics as well as execution timing, and periodically reports to the cloud server. 
The cloud server verifies if all ingested data is processed according to the pipeline (correctness), and if the edge incurs low delay (freshness).  

\paragraph{Thwarted attacks}
\sys{} defeats the following attacks.
\textbfit{i) Breaking IoT data confidentiality or integrity.}
As the raw and derived data enters and leaves the edge TEE through trusted IOs,
adversaries on the edge cannot touch, drop, or inject data. 
When the data is off the edge transmitted over untrusted networks,
it is protected by encryption against network-level adversaries. 
ii) \textbfit{Breaking the data plane integrity.}
Any fabricated opaque reference passed to the data plane will be rejected, since all opaque references are validated before use. 
Through the data plane's interface, an adversary may exploit bugs in the data plane and compromise it. 
By minimizing the date plane codebase and hardening its interface, \sys{} substantially reduces the data plane's attack surface and potential bugs that can be exploited. 
iii) \textbfit{Breaking analytics correctness.}
A compromised control plane may request computations
deviating from pipeline declarations or the stream model.
For instance, it may invoke trusted computations on partial data, wrong windows, or valid but undesirable opaque references. 
\sys{} defeats these attacks through attestation: since the cloud verifier possesses complete knowledge on ingested data and pipelines, it detects such correctness violation and rejects the edge analytics results.
\textbfit{iv) Attacks on analytics performance or availability.}
A compromised control plane may delay or pause invoking of trusted computations, violating the freshness guarantee. 
As the execution timing of trusted computations is attested, the cloud verifier detects the attacks and can choose to prompt further investigation.
\textbfit{v) Attempting to trigger data race or deadlock.}
By design, data race and deadlocks will never happen inside the data plane: 
the trusted computations do not share state concurrently and all locking happens outside of the TEE.

\section{Design Overview}
\label{sec:overview}




\subsection{Challenges}
\label{sec:overview:challenges}

Our approach raises three challenges.
i) \textbfit{Architecting the engine with a proper protection boundary}. 
This hinges on a key trade-off among TEE functional richness, overhead of TEE entry/exit, and TCB size. 
%
ii) \textbfit{Optimizing data functions within a TEE}. 
Processing of high-velocity data in a TEE strongly favors simple algorithms and compact memory. 
Yet, existing stream engines often operate numerous short-lived objects indexed in hash tables or trees~\cite{sparkstreaming,streambox,sensorbee,trill,streamscope}, e.g. for grouping events by key. 
They manage these objects with generic memory allocators~\cite{streambox} or garbage collectors~\cite{facade,sparkstreaming}. 
Such designs poorly fit a TEE's small TCB and limited DRAM portion, e.g. 
typically tens of MB for \TZ{} TEE and up to 128 MB for Intel SGX enclave ~\cite{secureKeeper}.
iii) \textbfit{Verifying stream analytics results}. 
This requires to track unbounded data flows in stream pipelines, 
validate if operators respect the temporal properties, e.g. windows, and minimize the resultant overhead in execution and communication.



\paragraph{Why are existing systems inadequate?} 
First,
many TEE-based systems~\cite{scone,graphene-sgx,haven} pull entire user applications 
and libraries to the TCB, as shown in Figure~\ref{fig:div}(a).
However, as we described in \sect{bkgnd:engine}, a modern analytics engine and its libraries are large, complex, and potentially vulnerable. 
Second,
partitioning applications to suit a TEE, as shown in Figure~\ref{fig:div}(b)~\cite{glamdring,panoply,rubinov16icse}, is unsuitable for existing stream engines: partitioning does not change their hash-based data structures and algorithms, which by design mismatch a TEE. 
Similarly, recent secure processing engines disfavor partitioning~\cite{enclaveDB,safebricks}.
Third,
recent systems use TEE to protect data in analytics or in network packet processing.
As summarized in Table~\ref{tab:cmp},
they lack support for stream analytics, key computation optimizations, or specialized memory allocation, which we will demonstrate as vital to our objective.


Attesting TEE integrity~\cite{sgx-tor,enclaveDB} is insufficient to assert analytics correctness.
VC3~\cite{vc3} and Opaque~\cite{opaque} verify correctness of \textit{batch} analytics by checking the history of compute results, i.e. their data lineage~\cite{data-provenance,ariadne}.
Without tracking data being continuously ingested and lacking a stream model, data lineages cannot assert whether \textit{all} ingested data is processed according to pipeline declarations, watermarks, and temporal windows, which are critical to stream analytics.


\subsection{\sysfull{} in a Nutshell}
\label{sec:overview:nutshell}

\sys{} builds on TrustZone~\cite{arm-trustzone} due to ARM's popularity for the edge and trusted IO benefiting stream analytics (\S\ref{sec:bkgnd}). 

\paragraph{Programmability}
Programming \sys{} is similar to programming commodity engines such as Spark Streaming~\cite{sparkstreaming} and Flink~\cite{apache-flink}.
Analytics programmers assemble pipelines with high-level, declarative operators as exemplified in Figure~\ref{fig:pipeline-stacked}(c). 
\sys{} provides most of the common operators offered by commodity engines, as summarized in Table~\ref{tab:operators}.
These stream operators are widely used for analytics over telemetry data~\cite{sensorbee,smart-power-plugs}.
\sys{} supports User Defined Functions (UDFs) that are certified by a trusted party, which is a common requirement in TEE-based systems~\cite{enclaveDB}. 

\paragraph{\sys{} architecture}
As shown in Figure~\ref{fig:arch},
\sys{}'s data plane incarnates as a \TZ{} module. 
\sys{} runs its control plane as a parallel runtime in the \nw{}. 
The control plane invokes the data plane through a narrow interface (details in \sect{eval}). 
The control plane orchestrates the execution of analytics pipelines. 
It creates plentiful parallelism among and within operators. 
It elastically maps the parallelism to a pool of threads it maintains.
At a given moment, all threads may simultaneously execute one operator as well as different operators over different data.

\paragraph{Data plane \& design choices}
\sys{}'s data plane consists of only the \task{}s and a runtime for them.


i) \Task{}s are stateless, single-threaded functions that are oblivious to synchronization.
We do not enclose whole stream pipelines in the data plane, because a stream pipeline must be scheduled dynamically for parallelism and handling high-velocity data.
We do not enclose whole declarative operators in the data plane, because
one operator instance has internal thread-level parallelism and hence requires thread management logic. 
Our choice keeps the data plane lean, leaving out all control functions including scheduling and threading.
This contrasts to many other engines pulling whole analytics to TEE as shown in Table~\ref{tab:cmp}.

Although exporting low-level primitives entails more TEE switches, 
the costs are lower on modern ARM~\cite{azab14ccs,vtz} and can be amortized by data batching, as will be discussed soon.


ii) The data plane incorporates minimum runtime functions:
memory management and paging, which are critical to TEE integrity;
cache coherence of parallel \tas{}s, which is critical to parallelism.
The data plane is agnostic to declarative operators and pipelines being executed. 


For attestation, the data plane generates audit records on data ingress/egress, watermarks, and \tas{} executions. 
It reduces overhead via data batching and record compression.

\paragraph{Coping with secure memory shortage}
When compute cost or data ingestion rate is high, 
\sys{} may run short of secure memory.
To avoid data loss in such a situation, \sys{} adds backpressure to source sensors, slowing down data ingestion.
In the current implementation, \sys{} triggers backpressure when ingestion exceeds a  user-defined threshold; we leave as future work automatic flow control, i.e. tuning the threshold online per available secure memory and backlog.

\section{\TASK{}s and Optimizations}
\label{sec:primitive}
\label{sec:opt}

\paragraph{Parallel execution inside a TEE}
\sys{} exploits task parallelism without bloating the TEE with a threading library.
The control plane invokes multiple \tas{}s from multiple worker threads, which then enter the TEE to execute the \tas{}s in parallel. 
All \task{}s share one cache-coherent memory address space in TEE, which simplifies data sharing and avoids copy cost. 
This contrasts to existing secure analytics engines that leave task parallelism untapped in a single TEE~\cite{vc3,securestreams}. 

\paragraph{Array-based algorithms to suit TEE}
Unlike many popular stream engines using hash-based algorithms for lower algorithmic complexity, we make a new design decision.
We strongly favor algorithms with simple logic and low memory overhead, despite that they may incur higher algorithmic complexity. 
Corresponding to contiguous arrays as the universal data containers in TEE, 
most \tas{}s use sequential-access algorithms over contiguous arrays, e.g. executing Merge-Sort over event arrays and scanning the resultant array to calculate the average value per key. 


\paragraph{\Task{}s and vectorization}
\sys{}'s \task{}s are generic. They constitute most declarative stream operators, often referred to as Select-Projection-Join-GroupBy (SPJG) families, shown in Table~\ref{tab:operators}. 
These operators are considered representative in prior research~\cite{mondrian}.

To speed up the array-based algorithms inside TEE without TCB bloat, 
our insight is to map their internal data parallelism to vector instructions of ARM~\cite{arm-neon}.
Despite their well-known performance benefit, vector instructions are rarely used to accelerate data analytics \textit{within} TEEs, to our knowledge. 
Vectorization incurs low code complexity as the performance gain comes from a CPU feature that is already part of the TCB. 


Our optimization focuses on Sort and Merge, two core primitives that dominate the execution of stream analytics according to our observation. 
Inspired by vectorized sort and merge on x86 ~\cite{sort-vs-hash-2009,sort-vs-hash-2013}, we build new implementations for \sys{} by hand-writing ARMv8 NEON vector instructions. 
Our sort outperforms the ones in the C/C++ standard libraries by more than 2$\times$, as will be shown in evaluation.
This optimization is crucial to the overall engine performance.


\section{TEE Memory Management}
\label{sec:mm}

Facing high-velocity streams in a TEE, 
\sys{}'s memory allocator addresses two challenges:
\textit{space efficiency}: it must create compact memory layout and reclaim memory timely due to limited physical memory;
\textit{lightweight}: the allocator must be simple to suit a low TCB. 	
The challenges disqualify popular engines that organize events in hash tables (e.g. for grouping events by key) and rely on generic memory allocators~\cite{sparkstreaming,streambox,sensorbee,trill,streamscope}.
The reasons are two: a hash table's principle of trading space for time mismatches TEE's limited memory; 
generic allocators often feature sophisticated optimizations, adding tens of KSLoC to TCB ~\cite{intel-tbb,jemalloc}.

\sys{} specializes memory management for stream computations: 
it supports unbounded buffers as the universal memory abstraction (\S\ref{sec:mm:ubuf}); 
it places data by using (untrusted) consumption hints and large virtual address space (\S\ref{sec:mm:placement}). 



\subsection{Unbounded Array}
\label{sec:mm:abs}
\label{sec:mm:ubuf}

%


We devise contiguous, virtually unbounded arrays called \textit{\ubuf{}}s, a new abstraction as the universal data containers used by computations in TEE. 
\ubuf{}s encapsulate all the data in a pipeline, including data flowing among \task{}s as well as operator states traditionally kept in hash tables.  

An \ubuf{} is an append-only buffer in a contiguous memory region for same-type data objects. 
Their lifecycles closely map to the producer/consumer pattern in streaming computations. 
One \ubuf{} can be in three states.
\textit{Open}: after an \ubuf{} is created, it dynamically grows as the producer \tas{} appends data objects to it.
\textit{Produced}: the data production completes and the end position of the \ubuf{} is finalized. \ubuf{} becomes read-only and no data can be appended.
\textit{Retired}: the \ubuf{} is no longer needed and its memory is subject to reclamation.
The memory allocator places and reclaims \ubuf{}s regarding their states, as will be discussed in Section~\ref{sec:mm:placement}.

\paragraph{Types}
\ubuf{}s fall into different types depending on their scopes and enclosed data.
A \textit{streaming \ubuf{}} encapsulates data flowing from a producer \tas{} to a consumer \tas{}. 
A \textit{state \ubuf{}} encapsulates operator state that outlives the lifespans of individual \tas{}s. 
A \textit{temporary \ubuf{}} live within a \task{}'s scope. 

\paragraph{Low abstraction overhead}
An \ubuf{} spans a contiguous virtual memory region and grows transparently. 
The growth is backed by the data plane's on-demand paging that completely happens in the TEE. 
For most of the time, growing an \ubuf{} only requires updating an integer index.
Compared to manually managed buffers, this mechanism waives bounds checking of \ubuf{} in computation code and hence allows the compiler to generate more compact loops. 
\ubuf{}s always grow in place.
This contrasts to common sequence containers (e.g. C++ \code{std::vector} and \code{java.util.ArrayList}) that grow transparently but require expensive relocation.
We will experimentally compare \ubuf{} with \code{std::vector} in Section~\ref{sec:eval}.

\subsection{Placing \ubuf{}s in \bufgroup{}s}
\label{sec:mm:placement}



\begin{wrapfigure}{}{0.20\textwidth}
\vspace{-4mm}
\includegraphics[width=0.20\textwidth,right]{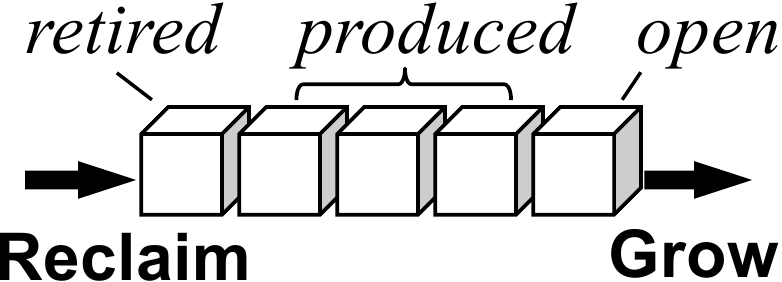}
\vspace{-6mm}		
\caption{The \ubuf{}s in one \bufgroup{}}
\label{fig:uarray}
\vspace{-15pt}		
\end{wrapfigure}

\paragraph{Co-locating \ubuf{}s}
The memory allocator co-locates multiple \ubuf{}s as a \bufgroup{} in order to reclaim them consecutively.
Spanning a contiguous virtual memory region, a \bufgroup{} consists of multiple \textit{produced} or \textit{retired} \ubuf{}s and optionally an \textit{open} \ubuf{} at its end, as shown in Figure~\ref{fig:uarray}.
The grouping is purely physical: 
it is at the discretion of the allocator, orthogonal to stream computations, and therefore transparent to the \task{}s and the control plane.




\setlength{\columnsep}{8pt}%

With the grouping, the allocator reclaims \textit{consumed} \ubuf{}s by always starting from the beginning of an \bufgroup{}, as shown in Figure~\ref{fig:uarray}.
To place a new \ubuf{}, the allocator decides whether to create a new \bufgroup{} for the \ubuf{}, or append the \ubuf{} to an existing \bufgroup{}.
In doing so, the allocator seeks to 
i) ensure that each \bufgroup{} holds a sequence of \ubuf{}s to be consumed consecutively in the future;
ii) minimize the total number of live \bufgroup{}s, in order to compact TEE memory layout and minimizes the cost in tracking \bufgroup{}s.
To this end, our key is to guide placement with the control plane's data consumption plan, as will be presented below. 






\paragraph{Consumption hints}
Upon invoking a \task{} $T$, the control plane may provide two optional hints concerning 
the future consumption order for the output of $T$: 

\begin{myitemize}
\vspace{-1mm}
\item 
\textit{Consumed-in-parallel} ($\parallel_k$): the control plane will schedule $k$ worker threads to consume a set of \ubuf{}s in parallel.
\vspace{-1mm}
\item
\textit{Consumed-after} ($b_1{\Leftarrow}b_2$): the control plane will schedule worker threads for consuming \ubuf{} $b_2$ after \ubuf{} $b_1$. 
The \textit{consumed-after} relation is transitive.
\ubuf{}s may form multiple \textit{consumed-after} chains.
\vspace{-1mm}
\end{myitemize}

The control plane may specify these relations between new output \ubuf{}s (yet to be created) and existing \ubuf{}s.
\paragraph{Hint-guided placement}
The hints assist the data plane to generate compact memory layout and reclaim memory effectively.
Upon allocating a \ubuf{}, the allocator examines the existing hints regarding to the \ubuf{}.

\noindent
($\Leftarrow$) prompts the allocator to place the \ubuf{}s on the same \textit{consumed-after} chain in the same \bufgroup{}. 
Starting from the new \ubuf{} $b$ under question,
the allocator tracks back on its consumed-after chain, and places $b$ after the first \ubuf{} that is both in state \textit{produced} (i.e. its growth has finished) and is located at the end of an \bufgroup{}.
If no such \ubuf{} is available on the chain, the allocator creates a new \bufgroup{} for $b$.

\noindent
($\parallel_k$) prompts the allocator to place \ubuf{}s $b_{1..k}$ in separate \bufgroup{}s, so that delay in consuming any of the \ubuf{}s will not block the allocator from reclaiming the other \ubuf{}s. 
Our rationale is that despite $b_{1..k}$ are created at the same time, they are often consumed at different moments in the future:
i) since \sys{}'s control plane threads independently fetch new \ubuf{}s for processing as they become available (\S\ref{sec:overview}), 
the starting moments for processing $b_{1..k}$ may vary widely, especially when the engine load is high;
ii) even when $k$ worker threads start processing $b_{1..k}$ simultaneously, straggling workers are not uncommon, due to non-determinism of a modern multicore's thread scheduling and memory hierarchy~\cite{Aviram:2012:ESD:2160718.2160742}.

\paragraph{The impacts of misleading hints}
\sys{} detects misleading hints in retrospect through remote attestation (\S\ref{sec:attest}).
As the hints only affect TEE memory placement \textit{policy} on the edge,
misleading hints never result in data loss (\S\ref{sec:overview:nutshell}) or violation of data security and TEE integrity.
Yet, such hints may slow down analytics and therefore violate result freshness.

\paragraph{Managing virtual addresses}
All \bufgroup{}s grow \textit{in place} within one virtual address space. 
To avoid collision and expensive relocation, the allocator places them far apart by leveraging the large virtual address space dedicated to a \TZ{} TEE. 
The space is 256TB on ARMv8, 10,000$\times$ larger than the physical DRAM (a few GBs).
Hence, the allocator simply reserves for each \bufgroup{} a virtual address range as large as the total TEE DRAM. We will validate this choice in \sect{eval}.
\newcommand{\dfrom}{\textit{derived-from}}

\section{Attestation for Correctness and Freshness}
\label{sec:attest}

\sys{} collects evidences for cloud consumers to verify two properties:
\textit{correctness}, i.e. all ingested data is processed according to the stream pipeline declaration;
\textit{freshness}, i.e. the pipeline has low output delays. 

The above objective has several notable aspects. 
i) We verify the behaviors of untrusted control plane, i.e., \textit{which} primitives it invokes on \textit{what} data and at \textit{what} time.
We do not verify \task{}s, e.g. if a Sort primitive indeed produces ordered data.
ii) 
Verifying data lineages at the pipeline's intermediate operators or egress~\cite{data-provenance,ariadne} 
is insufficient to guarantee correctness, i.e. all data ingested so far is processed according to the stream pipeline. 
iii) 
The windows of stream computations and watermarks triggering the computations must be attested, which are keys to stream model
(\S\ref{sec:bkgnd}).
iv) As the volume of evidences can be substantial, evidences must be compacted to save uplink bandwidth~\cite{farmbeats,satellites-iot}. 

Therefore, \sys{} provides the following verification mechanism. 
Agnostic to the pipeline being executed,
the data plane monitors dataflows among \tas{} instances at the TEE boundary, 
and then generates audit records. 
For low overhead, it eschews building data lineages on-the-fly unlike much prior work~\cite{ariadne,protracer,vc3}. 
The data plane compresses audit records and flushes to the cloud both periodically and upon externalizing any analytics result.
We describe details below. 





\begin{figure}[t!]
\centering
\vspace{3pt}
\includegraphics[width=0.42\textwidth{}]{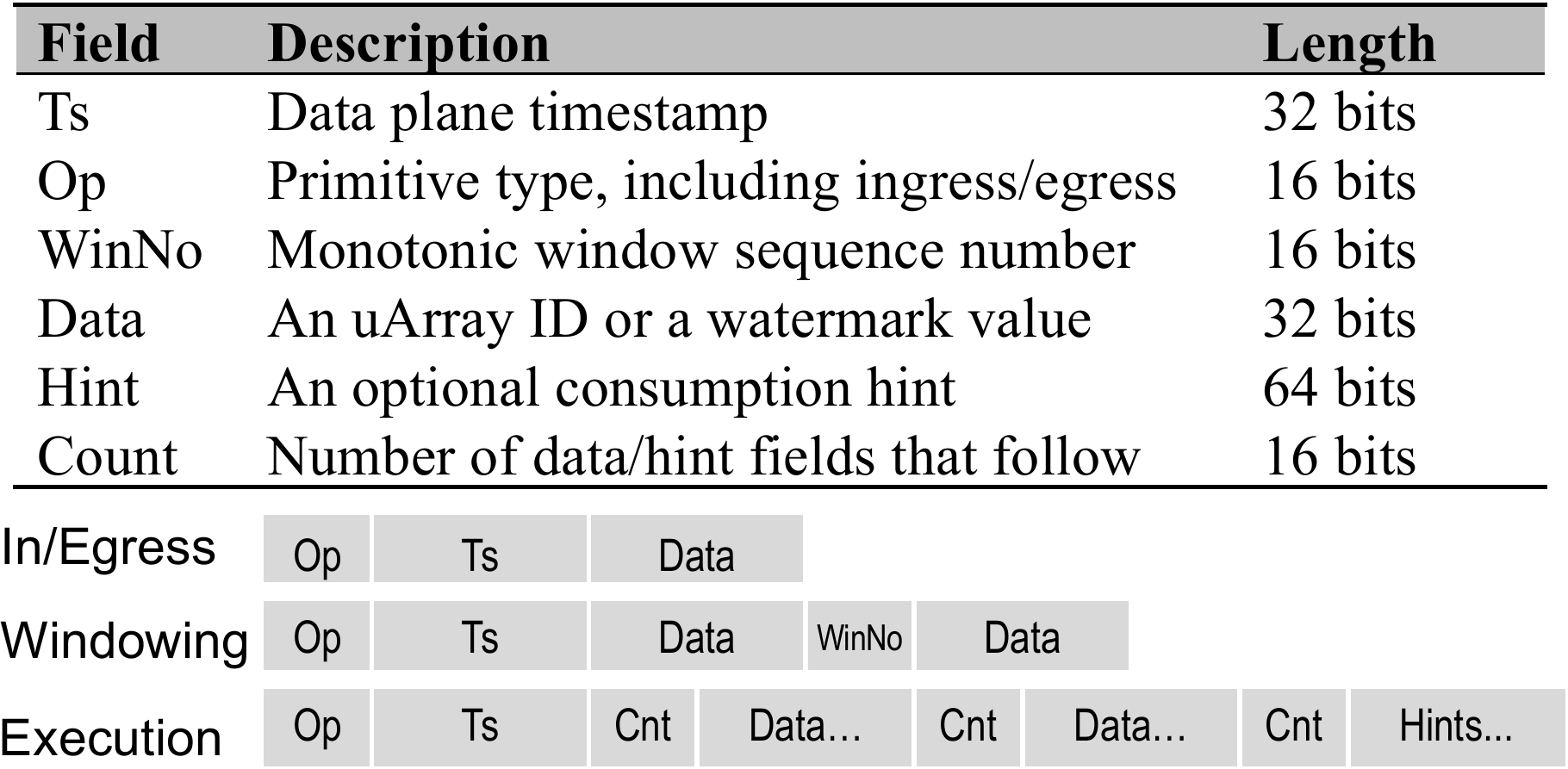} 
\vspace{-6pt}		
\caption{Audit records: fields (top) and layout (bottom)}
\label{fig:attest}
\vspace{-18pt}		
\end{figure}

\paragraph{Audit records} 
As being invoked by the control plane, 
the data plane generates \textit{audit records}. 
As illustrated in Figure~\ref{fig:attest}, the records track
i) ingested and externalized \ubuf{}s, ii) associations between \ubuf{}s and windows, 
and iii) \tas{} executions (with optional hints supplied by the control plane) which establish \dfrom{} relations among \ubuf{}s. 
The records further include ingested watermark values,
which are crucial for determining output delays as will be discussed below.
The data plane timestamps all the records.
It generates monotonically increasing identifiers for recorded \ubuf{}s. 
We will evaluate the overhead of audit records in \sect{eval}.

\paragraph{Attesting analytics correctness}
The cloud verifier checks if all ingested \ubuf{}s flow through the expected \task{}s. 
Such dataflows are deterministic given the arrivals of input data (including their windows), the watermarks, and the pipeline declaration. 
Hence, the verifier replays all ingestion records on its local copy of the same pipeline. 
It checks if all the records resulting from the replay match the ones reported by the edge (except timestamps). 
The replay is symbolic without actual computations and hence fast. 


Note that the verification works for stateful operators as well. 
The state of a stream operator (e.g. temporal join) is only determined by all the inputs the operator has ever received.
Since the cloud can verify that all the ingested \ubuf{}s correctly flow through the expected \task{}s and thus stream operators, it knows that the operator's current state must be correct, and then all results derived from the operator state must be correct.

\paragraph{Attesting result freshness}
The key for the verifier to calculate the delay of an output result $R$ is to identify the watermark that triggers the externalization of $R$, according to the delay definition in Section~\ref{sec:bkgnd:pipeline}.
From the egress record of $R$, the verifier traces \textit{backward} following the \dfrom{} chain(s) until it reaches an execution record indicating that a watermark $W$ triggers the execution. 
The verifier looks up the ingress record of $W$. 
It calculates the difference between $W$'s ingress time and $R$'s egress time to be the delay of $R$. 

\paragraph{Example}
In Listing~\ref{list:attest}, an uArray with identifier 0xF0 is ingested and segmented into two \ubuf{}s (0xF1 and 0xF2) for window 0 and 1 respectively. 
Sort consumes \ubuf{} 0xF1 and produces \ubuf{} 0xF3. 
A watermark with value 100 arrives and completes window 0.
Triggered by the watermark, SUM consumes \ubuf{} 0xF3 of window 0 and produces \ubuf{} 0xF5 as the result of window 0. 

\vspace{-1mm}
\lstset{
	language=C++,
	basicstyle=\fontsize{8}{8}\selectfont\ttfamily,
	frame=tb, 
	breaklines=true,
	captionpos=b,
	label=list:attest}
\begin{lstlisting}[caption={Sample audit records for the pipeline in Figure~\ref{fig:pipeline-stacked}. Format is simplified. ts means processing timestamp.}, escapechar=!]
ts= 1 INGRESS data=0xF0
ts= 5 WND data_in=0xF0 win_no=0 data_out=0xF1
ts=10 SORT data_in=0xF1 data_out=0xF3
ts=15 INGRESS data=0xF4 (watermark=100)
ts=25 SUM data_in=0xF3,0xF4 data_out=0xF5
ts=28 WND data_in=0xF0 win_no=1 data_out=0xF6 
ts=30 EGRESS data=0xF5
\end{lstlisting}
\vspace{-2mm}
The cloud verifier replays the ingress records on its local pipeline copy and learns that \ubuf{} 0xF1 is processed adhering to the pipeline declaration while \ubuf{} 0xF2 is yet to be processed. 
It will assert analytics incorrectness if 0xF2 remains unprocessed until a future watermark completes window 1 (not shown). 
To verify result freshness, the verifier traces result 0xF5 backward to find its trigger watermark 0xF4 and calculates the output delay to be 15 ($30-15$).

\paragraph{Columnar compression of records}
The data plane compresses audit records by exploiting locality within one record field and known data distribution in each field. 
The data plane produces raw audit records in memory (with the format shown in Figure~\ref{fig:attest}) and in a row order, i.e. one record after the other.
Before uploading a sequence of records, it separates the record fields (i.e. columns) and applies different encoding schemes to individual columns: 
i) Huffman encoding for primitive types and data counts, the two columns likely contain skewed values; 
ii) delta encoding for timestamps, \ubuf{} identifiers, and window numbers, which increment monotonically.
Our compression is inspired by columnar databases~\cite{c-store}. 
We will evaluate the efficacy of compression in \sect{eval}.


\section{Implementation}
\label{sec:impl}

We build \sys{} for ARMv8 and atop OP-TEE~\cite{optee} (v2.3).
\sys{} reuses most control functions of StreamBox~\cite{streambox}, an open-source research stream engine for x86 servers. 
Yet, as StreamBox mismatches a TEE (\S\ref{sec:overview:challenges}), \sys{} contributes a new architecture and a new data plane. 
\sys{} communicates with source sensors and cloud consumers over ZeroMQ TCP transport~\cite{zmq} which is known for good performance.
The new implementation of \sys{} includes 12.4K SLoC.

\noindent
\textbf{Input batch size}, a key parameter of \sys{}, trades off between delays in executing individual \tas{}s, the rate of TEE entry/exit, and attestation cost. 
We empirically determine it as 100K events and will evaluate its impact (\S\ref{sec:eval}).
%
\noindent
\textbf{Opaque references} for \ubuf{}s are 64-bit random integers generated by the data plane. 
It keeps the mappings from references to \ubuf{} addresses in a table, and 
validates opaque references by table lookup.
This incurs minor overhead, as live opaque references are often no more than a few thousands.

\section{Evaluation}
\label{sec:eval}


\begin{figure*}
    \centering
    \hspace{-3pt}
    \begin{subfigure}[b]{0.30\textwidth}
    	\includegraphics[width=\textwidth]{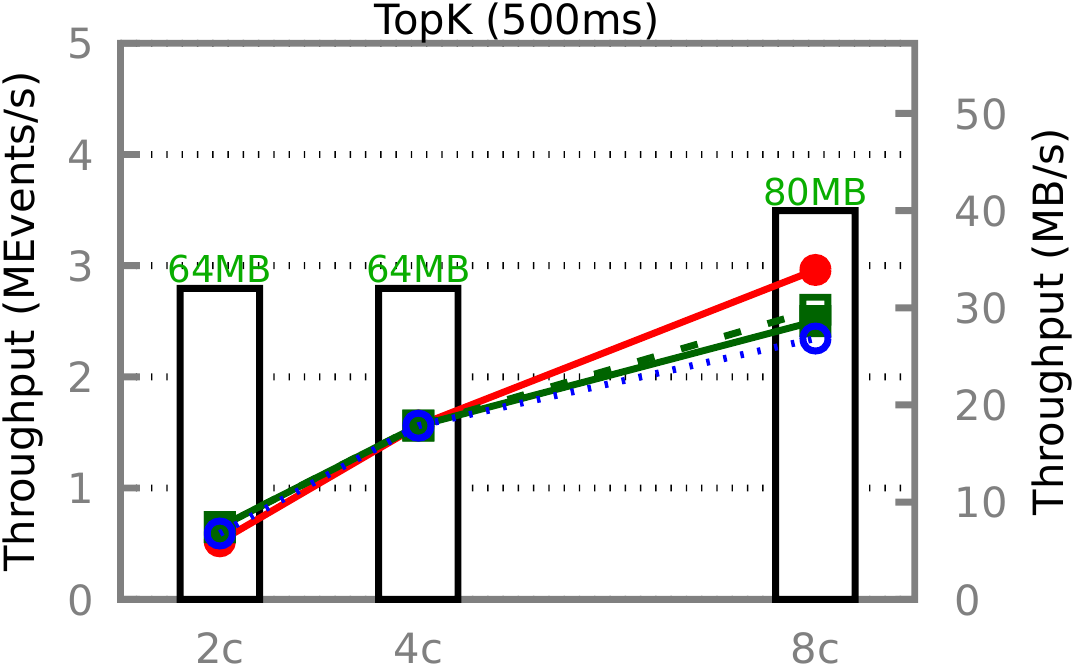}
		\label{fig:top-value}
        \vspace{-5pt}		
	\end{subfigure}
	~
    \hspace{7pt}
    \begin{subfigure}[b]{0.315\textwidth}
		\includegraphics[width=\textwidth]{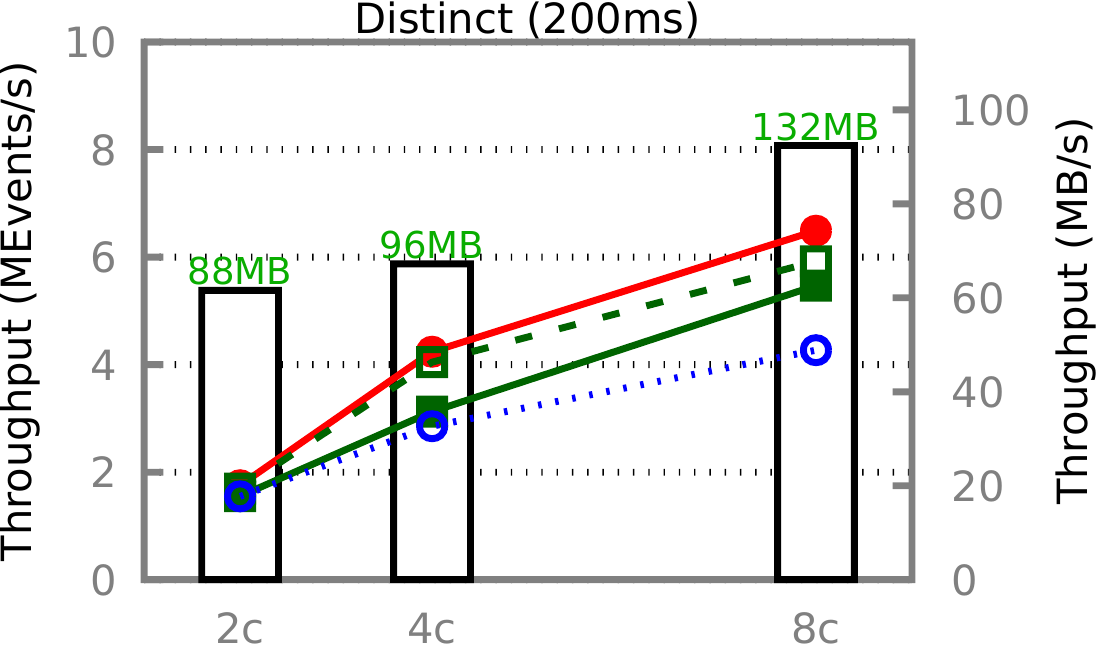}
        \label{fig:unique-taxis}
        \vspace{-5pt}		
	\end{subfigure}    
    ~
    \begin{subfigure}[b]{0.313\textwidth}
        \includegraphics[width=\textwidth]{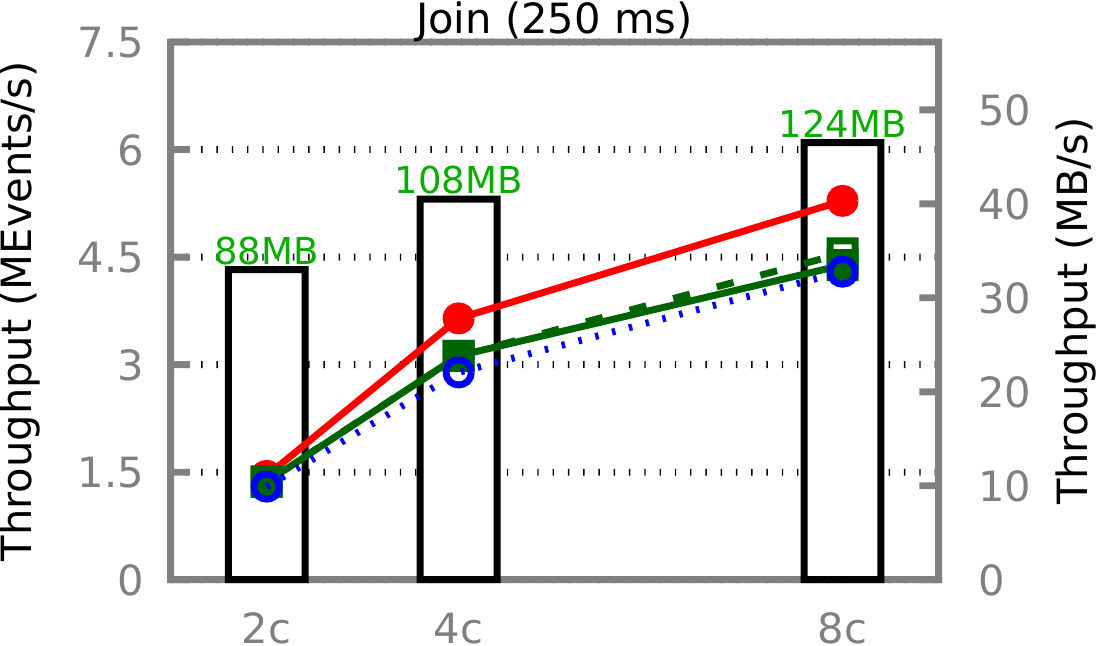}
        \label{fig:temp-join}
        \vspace{-5pt}		
    \end{subfigure}

	
	\begin{subfigure}[b]{0.315\textwidth}
	    \includegraphics[width=\textwidth]{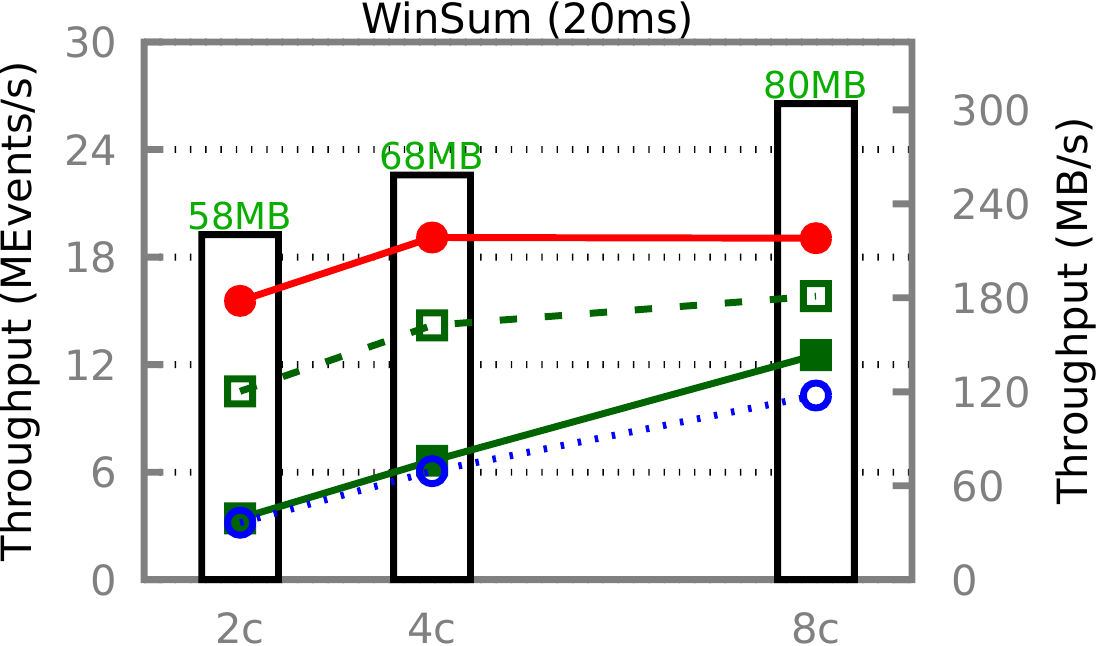}
		\label{fig:winsum}
	\end{subfigure}
	~
    \hspace{2.5pt}
	\begin{subfigure}[b]{0.315\textwidth}
		\includegraphics[width=\textwidth]{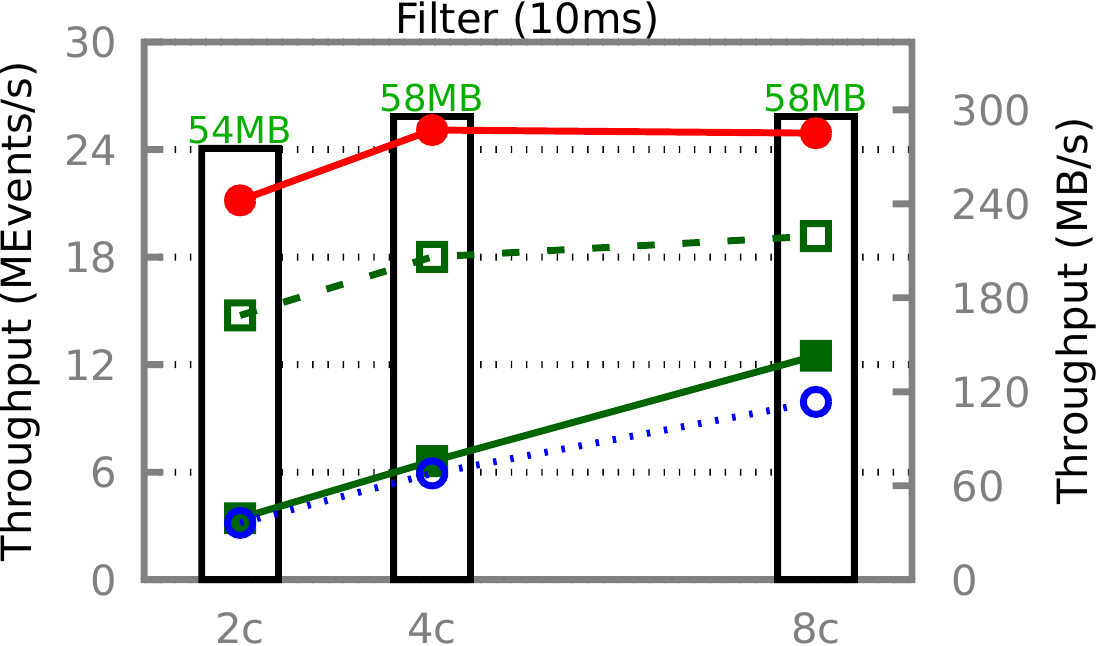}
		\label{fig:filter}
	\end{subfigure}    
    ~
	\hspace{2.5pt}
    \begin{subfigure}[b]{0.315\textwidth}
        \includegraphics[width=\textwidth]{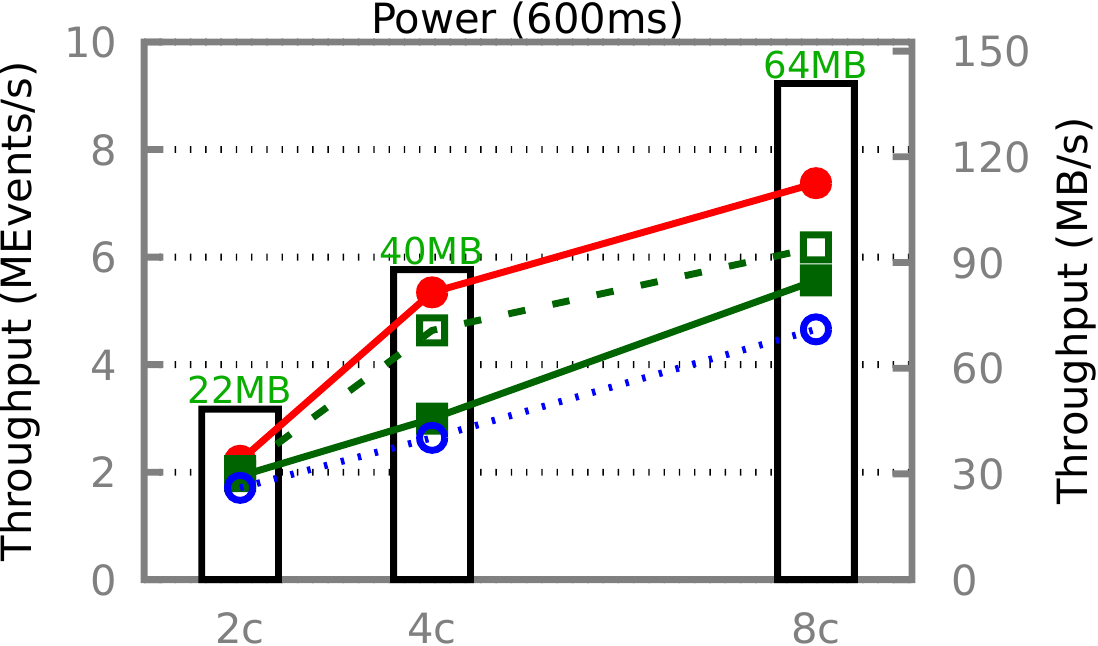}
        \label{fig:power-grid}
	\end{subfigure}

    \vspace{-15pt}		

    \caption{\sysfull{} throughput (lines, left/right y-axes) as a function of CPU cores (x-axis) under given output delays (above each plot). Steady consumptions of TEE memory as columns with annotated values.
    See Table~\ref{tab:exp} for legends and explanations.
      }
    \label{fig:scalability}
    
    \vspace{-14pt}		
    
\end{figure*}

%
%
%

We answer the following questions through evaluation:
\begin{smitemize}
\item Does \sys{} result in a small TCB? (\S\ref{sec:eval:source})
\vspace{-2pt}
\item What is \sys{}'s performance and how is it compared to other engines? What is the overhead? (\S\ref{sec:eval:perf})
\vspace{-2pt}
\item How do our key designs impact performance (\S\ref{sec:eval:feature})? 
\end{smitemize}

\subsection{TCB Analysis}
\label{sec:eval:source}


\paragraph{TCB size}
Table~\ref{tab:src} shows a breakdown of the \sys{} source code. 
Despite a sophisticated control plane, 
the data plane only adds 5K SLoC to the TCB. 
\sys{}'s memory management is in 740 SLoC, 9$\times$ fewer than glibc's \code{malloc} and 20$\times$ fewer than \code{jemalloc}~\cite{jemalloc}.
The size of data plane is 42.5~KB, a small fraction (16\%) of the entire OP-TEE binary.

\paragraph{TCB interface}
The \sys{}'s data plane exports only four entry functions:
two for data plane initialization/finalization, 
one for debugging,
and one shared by all 23 \task{}s.
The last function accepts and returns opaque references (\S\ref{sec:overview}).
No state is shared across the protection boundary. 

\paragraph{Comparison with alternative TCBs}
Compared to enclosing whole applications in TCB~\cite{scone,haven,graphene-sgx},
\sys{} keeps most of the engine out, shrinking the TCB by at least one order of magnitude. 
Compared to directly carving out~\cite{glamdring,rubinov16icse} the original StreamBox's data functions for protection, \sys{} completely avoids sophisticated data structures (e.g.  AtomicHashMap~\cite{folly} used by StreamBox)  that mismatch TCB.
Compared to VC3~\cite{vc3} that implements Map/Reduce operators in a TCB with $\sim$9K SLoC, \sys{} supports much richer stream operators within a 2$\times$ smaller TCB. 

\subsection{Performance \& Overhead}
\label{sec:eval:perf}


\begin{table}[t!]
\centering
\vspace{2pt}
\includegraphics[width=0.45\textwidth{}]{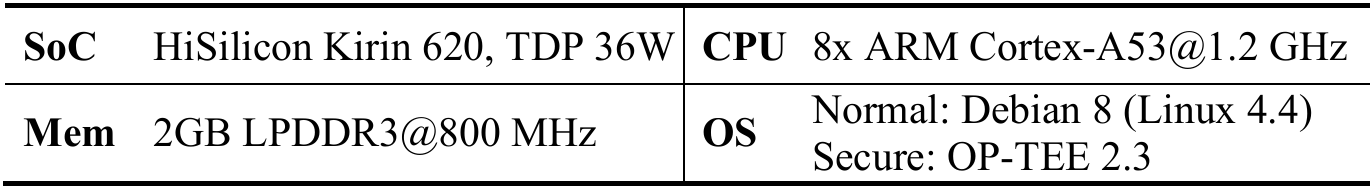}
\vspace{-10pt}
\caption{The test platform used in experiments}
\label{tab:plat}
\vspace{-2.5mm}		
\end{table}


\begin{table}[t!]
\centering
\includegraphics[width=0.45\textwidth{}]{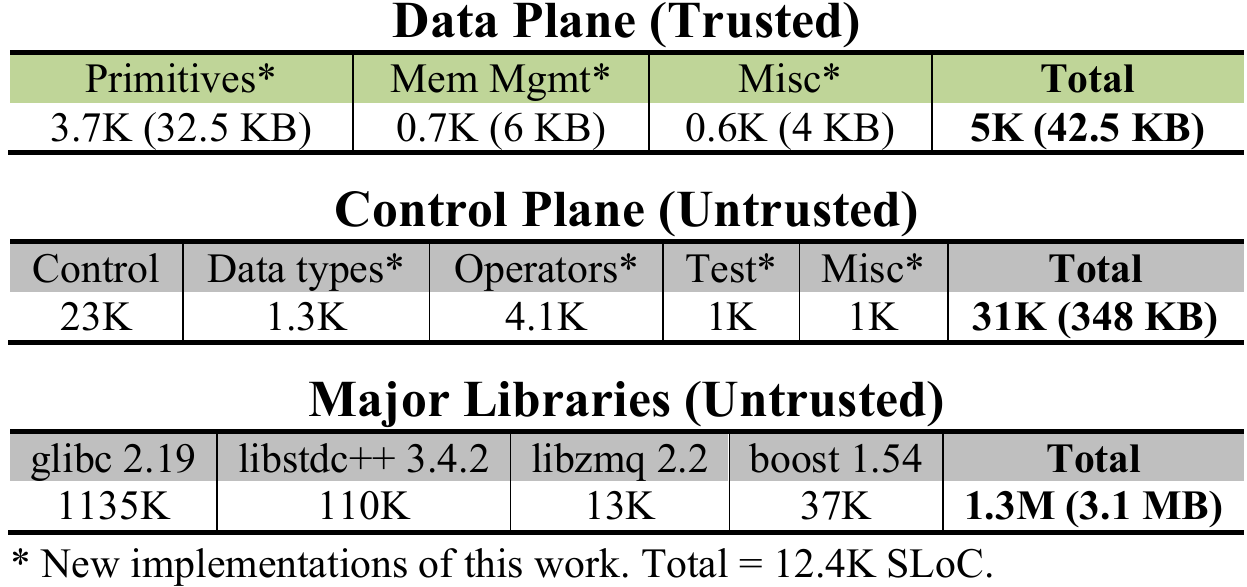} 
\vspace{-7pt}		
\caption{A breakdown of the \sysfull{} source, of which 5K SLoC are in TCB.
Binary code sizes shown in parentheses
}
\label{tab:src}
\vspace{-10pt}		
\end{table}
\paragraph{Methodology}
We evaluate \sys{} on a HiKey board as summarized in Table~\ref{tab:plat}.
We chose HiKey for its good OP-TEE support~\cite{optee} and that it is among the few boards with \TZ{} programmable by third parties.
We built \sender{}, a program sends data streams over ZeroMQ TCP transport~\cite{zmq} to \sys{}.
We run the cloud consumer on an x86 machine. 
Data streams are encrypted with 128-bit AES. 
%

In the face of HiKey's platform limitations, we set up the engine ingestion as follows. 
i) Although Gigabit Ethernet on edge platforms is common~\cite{marvell-armada-8k,imx7}, 
Hikey's Ethernet interface (over USB) only has 
20MB/sec bandwidth. 
We have verified that the interface is saturated by \sys{} with 4 cores. 
Hence, we report performance when \sys{} and \sender{} both run on HiKey communicating over ZeroMQ TCP, which still fully exercise the TCP/IP stack and data copy. 
ii)
Although HiKey's TEE is capable of directly operating Ethernet interface as trusted IO, our OP-TEE version lacks the needed drivers.
Hence, we emulate \sys{}'s direct data ingestion to TEE by running the ingestion in a privileged process in the \nw{}, and bypassing data copy across the TEE boundary.
Our test harness continuously replays pre-allocated secure memory buffers populated with events.

As summarized in Table~\ref{tab:exp}, we test \sys{} as well as three modified versions: 
\textbf{\sysdirclr{}} ingests data in cleartext; this is allowed if source-edge links are trusted as defined in our threat model (\S\ref{sec:bkgnd:secmodel}). 
\textbf{\sysindenc{}} does not exploit \TZ{}'s trusted IO: the untrusted OS ingested (encrypted) data and copies the data across TEE boundary to the data plane. 
\textbf{\unsecure{}} completely runs in the \nw{} with ingress and egress in cleartext, showing native performance. 
This is basically StreamBox~\cite{streambox} with \sys{}'s optimized stream computations (\S\ref{sec:primitive}). 
We report the engine performance as its maximum input throughput when the pipeline output delay (defined in \S\ref{sec:bkgnd:pipeline}) remains under a target set by us.

\begin{table}[]
\centering
\includegraphics[width=0.47\textwidth{}]{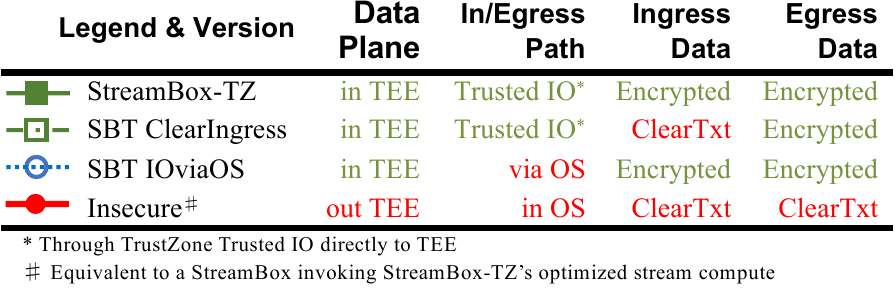}
\vspace{-8pt}		
\caption{Engine versions for comparison (plots in Figure~\ref{fig:scalability})}
\label{tab:exp}
\vspace{-18pt}		
\end{table}
\paragraph{Benchmarks}
We employ six benchmarks of processing sensor data streams from
prior work~\cite{DEBS15_taxi,smart-power-plugs,saber,trill,streambox}.
They cover major stream operators and a variety of pipelines. 
We use fixed windows, each encompassing 1M events and spanning 1 second of event time.
Each event consists of 3 fields (12 Bytes) unless stated otherwise.
(1)
\textbf{Top Values Per Key (\freq{})} groups events based on keys and identifies the K largest values in each group in each window.
(2)
\textbf{Counting Unique Taxis (\unique{})} identifies unique taxi IDs and counts them per window.
For input events, we use a dataset of taxi trip information containing 11~K distinct taxi IDs~\cite{DEBS15_taxi}.
(3)
\textbf{Temporal Join (\join{})} joins events that have the same keys and fall into same windows from two input streams.
(4)
\textbf{Windowed Aggregation (\winsum{})} aggregates input values within each window.
We use the Intel Lab Data~\cite{labdata_intel} consisting of real sensor values as input.
(5)
\textbf{Filtering (\filter{})} filters out input data, of which field falls into to a given range in each window. We set 1\% selectivity as done in prior work~\cite{saber}. 
(6)
\textbf{Power Grid (\grid{})}, derived from a public challenge~\cite{smart-power-plugs}, finds out houses with most high-power plugs.
Ingesting a stream of per-plug power samples, it calculates the average power of each plug in a window and the average power over all plugs in all houses in the window.
For each house, it counts the number of plugs that have a higher load than average. 
It emits the houses that have most high-power plugs in the window.
The event for this benchmark is composed of 4 fields (16 Bytes).
Benchmark 2, 4, and 6 use real-world datasets; 
others use synthetic data sets of which fields are 32-bit random integers. 
Note that \sys{}'s \code{GroupBy} operator bases on sort and merge and is insensitive to key skewness~\cite{Albutiu:2012:MPS:2336664.2336678}.

\begin{figure*}
    \centering
    \begin{minipage}[t]{0.32\textwidth}
            \includegraphics[width=\textwidth]{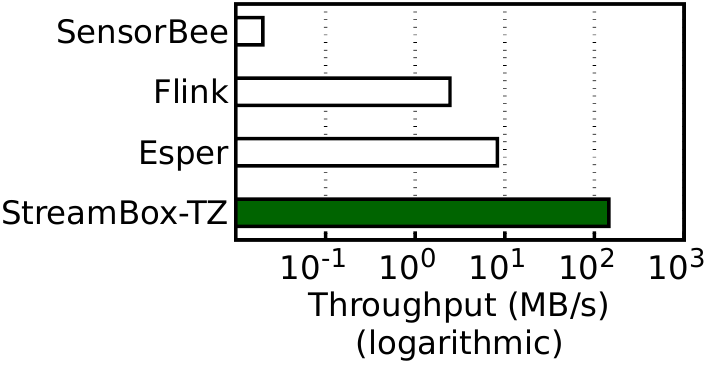}
	   	    \vspace{-12pt}		
			\vspace{-5pt}		
            \caption{\sysfull{} achieves much higher throughput than commodity insecure engines~\cite{sensorbee,apache-flink,esper} on HiKey. 
            Benchmark: windowed aggregation; target output delay: 50ms.            
            }
			\label{fig:esper}
	\end{minipage}
    ~
    \begin{minipage}[t]{0.32\textwidth}
        \includegraphics[width=\textwidth]{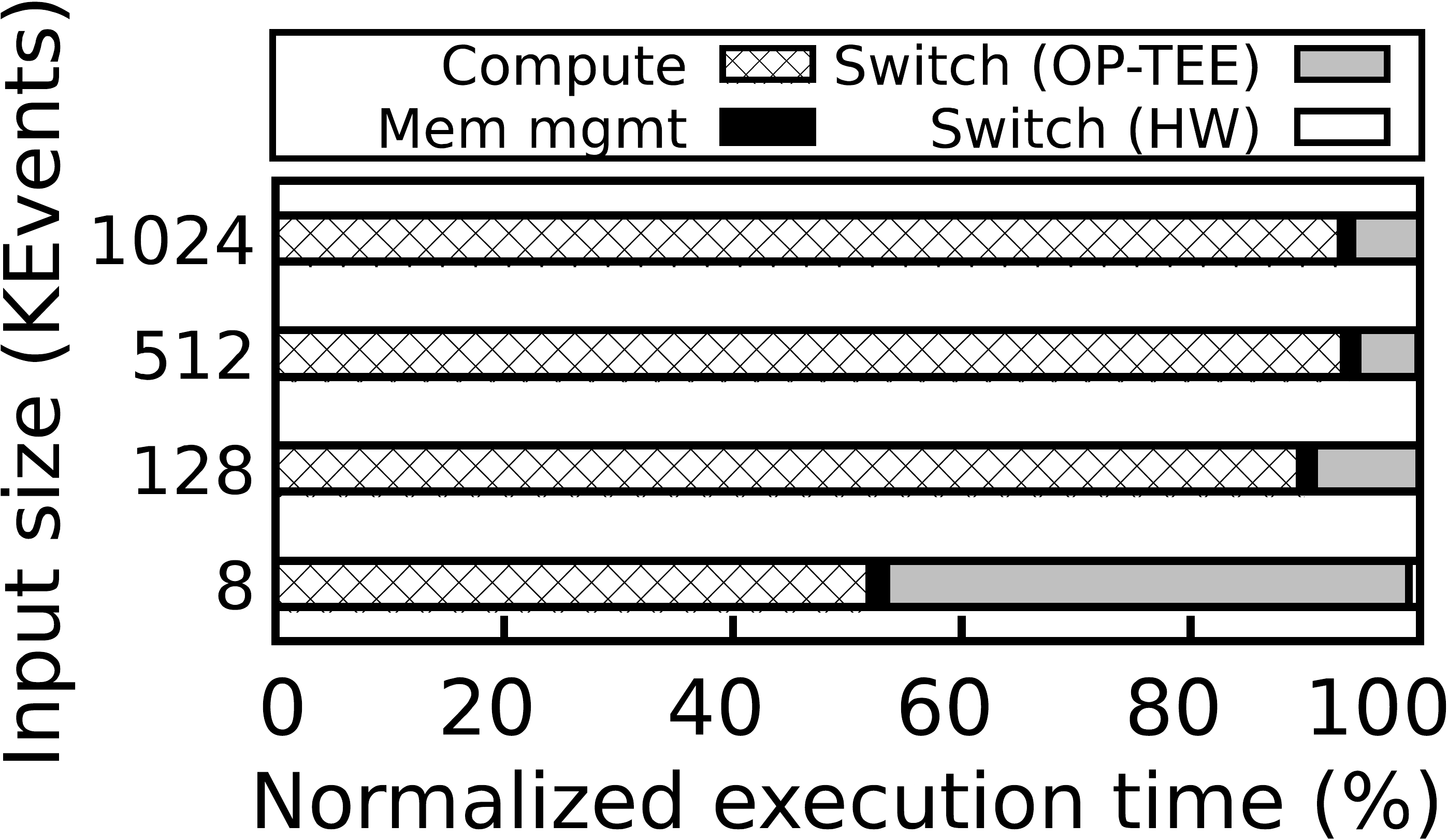}
   	    \vspace{-12pt}		
   	    \vspace{-5pt}		
        \caption{Run time breakdown of operator \code{GroupBy} under different input batch sizes.
        	The control plane runs 8 threads to execute \code{GroupBy} in parallel.   
        	Total execution time is normalized to 100\%.
        	}
		\label{fig:overhead}
    \end{minipage}
    ~
	\begin{minipage}[t]{0.32\textwidth}
    	\includegraphics[width=\textwidth]{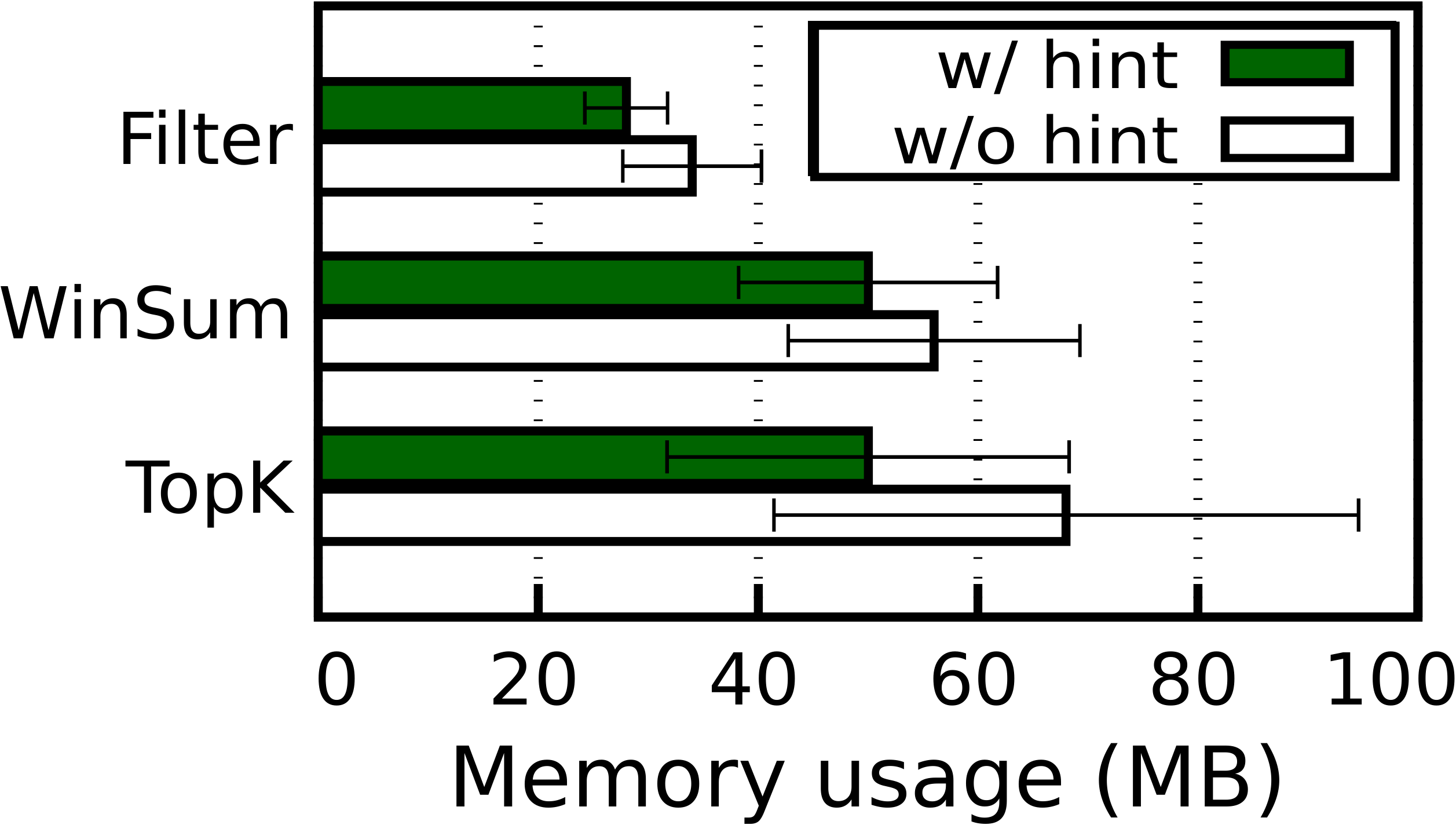}
    	\label{fig:cmp_sort}
   	    \vspace{-12pt}		
   	    \vspace{-5pt}		
        \caption{Without consumption hints, the allocator uses more TEE memory. Since memory usage fluctuates at run time, the error bars show two standard deviations below and above the average.}
		\label{fig:hint}
	\end{minipage}    
    \vspace{-14pt}		
    
\end{figure*}
\paragraph{End-to-end performance}
Figure~\ref{fig:scalability} shows the throughputs of all benchmarks as a function of hardware parallelism.
\sys{} can process up to multiple millions of events within sub-second output delays (labeled atop each plot).
For simpler pipelines such as \winsum{} and \filter{}, \sys{} processes around 12M events/sec (140 MB/sec). 
This throughput saturates one GbE link which is common on IoT gateways
~\cite{imx7}.
%
Overall, \sys{} can use all 8 cores in a scalable manner. 

\sys{}'s absolute performance is state of the art. 
We test three popular, insecure stream engines: 
Flink~\cite{apache-flink}, designed for distributed environment and known for good single-node performance~\cite{flink-single-perf};
Esper~\cite{esper}, designed for a single machine; 
SensorBee~\cite{sensorbee}, designed for sensor data processing on a single device.
As shown in Figure~\ref{fig:esper}, on the same hardware (HiKey) and the same benchmark (\winsum{}),
we have measured that \sys{}'s throughput is at least one order of magnitude higher than the others.
This is because i) our \textit{Insecure} baseline has high performance for its rich task parallelism (inherited from StreamBox~\cite{streambox}) and native, vectorized stream computations (new contributions);
ii) \sys{} only imposes modest security overhead, as will be shown later.

\paragraph{Comparison to secure stream engines}
The comparison is challenged by that TrustZone was rarely exploited for protecting data-intensive computations. 
To our knowledge, i) no analytics engines use TrustZone for data protection and ii) no systems can partition an insecure stream engine for TrustZone.
Note that popular secure analytics engines, e.g. VC3~\cite{vc3} and Opaque~\cite{opaque}, not only require SGX but also target batch processing instead of stream analytics.
To this end, we qualitatively compare \sys{} with SecureStreams~\cite{securestreams}, the closest system we are aware of.
Designed for an x86 cluster, SecureStreams uses SGX to protect stream operators and targets strong data security.
On a benchmark similar to WinSum it was reported to achieve 10 MB/sec, one magnitude lower than \sys{} on WinSum.
Furthermore,
SecureStreams achieved such performance on a small x86 cluster which has much richer resource than HiKey: the former has 
faster CPUs (8x i7-6700@3.4GHz versus 8x Cortex-A53@1.2GHz), larger DRAM (16~GB versus 2~GB), higher power (130W versus 36W), and higher cost (\$600 versus \$65).

\sys{}'s advantage comes from i) data exchange via coherent memory inside one TEE, instead of exchanging encrypted messages among workers; ii) memory management specialized for streaming, and iii) vectorized computations.
\paragraph{Security overhead}
We investigate the overhead of the new security mechanism contributed by \sys{} -- its isolated data plane. 
We assess the overhead as the throughput loss of \sysdirclr{} as compared to \unsecure{} (i.e. native performance as StreamBox~\cite{streambox} invoking \sys{}'s stream computations), both paying same costs for data ingress.
The target output delays are the same (labeled atop each plot in Figure~\ref{fig:scalability}). 
The security overhead is less than 25\% in all benchmarks.
This is similar to or lower than the reported overhead (20--70\%) in recent TEE systems~\cite{glamdring,graphene-sgx,scone}.
\textit{\textbf{Overhead analysis:}}
The security overhead mostly comes from world switch, among operators and inside each operator.
To understand the switch cost within an operator, 
we profile \code{GroupBy}, one of the most costly operators. 
We test different input batch sizes, which have a strong impact on TEE entry/exit rates and hence isolation overhead (\S\ref{sec:overview}).
Figure~\ref{fig:overhead} shows a run time breakdown. 
When each input batch contains 128K (close to the value we set for \sys{}) or more events, 
more than 90\% of the CPU time is spent on actual computations in TEE. 
The CPU usage of TEE memory management is as low as 1--2\%.
In the extreme case where each input batch contains as few as 8K events, the overhead of world switch starts to dominate. 
Most of the world switch overhead comes from OP-TEE instead of the CPU hardware (a few thousand cycles per switch), 
suggesting room for OP-TEE optimization. 





\paragraph{Impact of decrypting ingress data}
Decrypting ingress data
is needed 
if source-edge links are untrusted (\S\ref{sec:bkgnd:secmodel}) and source must send encrypted data.
It has substantial performance impact. 
By comparing \sys{} to \sysdirclr{}, 
turning on/off ingress decryption leads to 4\% -- 35\% throughput difference when all 8 cores are in use.
The performance gap is more pronounced for simple pipelines, which has higher ingestion throughput leading to  higher decryption cost. 

\paragraph{TEE memory usage}
While sustaining high throughput, \sys{} consumes a moderate amount of physical memory, ranging from 20~MB to 130~MB as shown in Figure~\ref{fig:scalability}.
The memory usage is as low as 1--6\% of the total system DRAM. 
The virtual memory usage is also low, 
often 1--5\% of the entire virtual address space in OP-TEE.
The memory usage increases with the throughput, since there will be more in-flight data. 
On the same platform, Flink's memory consumption is 3$\times$ higher, due to its hash-based data structures and the use of JVM. 
This validates our choice of \ubuf{}s. 




\paragraph{Attestation overhead}
Attestation incurs minor overhead to both the edge and the cloud. 
We measured that \sys{} produces 300--400 audit records per second across all our benchmarks, and spends a few hundred cycles on producing each record. 
Compressing such record streams on HiKey consumes 0.2\% of total CPU time.
Our consumer written in Python on a 4-core i7-4790 machine replays 57K records per second with a single core, 
suggesting a capability of attesting near 500 \sys{} instances simultaneously. 
We will evaluate the efficacy of record compression in Section~\ref{sec:eval:feature}.

%

\subsection{Validation of Key Design Features}
\label{sec:eval:feature}

\paragraph{Exploitation of trusted IO}
As shown in Figure~\ref{fig:scalability}, a comparison between \sys{} and \sysindenc{} demonstrates the advantage of directly ingesting data into TEE and bypassing the OS:
\sys{} outperforms the latter by up to 20\% in throughput due to reduction in moving ingested data. 


\paragraph{\Task{} vectorization} (\S\ref{sec:opt})
Our optimizations with ARM vector instructions are crucial. 
To show this, we examine \code{GroupBy}, one of the top hotspot operators. 
When we replace the vectorized Sort that underpins \code{GroupBy} with two popular implementations (qsort() from the the OP-TEE's libc and std::sort() from the standard C++ library), 
we measured the throughput of \code{GroupBy} drops by up to 7$\times$ and 2$\times$, respectively.
We have similar observation on other operators. 

\paragraph{Efficacy of hint-guided memory placement} (\S\ref{sec:mm:placement})
We compare to an alternative design: 
the modified allocator acts based on the heuristics that all the \ubuf{}s produced by the same \tas{} belong to the same \textit{generation} and are likely to be reclaimed altogether. 
Accordingly, the modified allocator places these \ubuf{}s in the same \bufgroup{}. 
As shown in Figure~\ref{fig:hint}, in three benchmarks, the modified allocator increases memory usage by up to 35\%.
This is because, without hints, it cannot place \ubuf{}s based on future consumption.

\begin{figure}
\centering
    \begin{minipage}[]{0.135\textwidth}
    \vspace{3pt}		
    \includegraphics[width=1.0\textwidth]{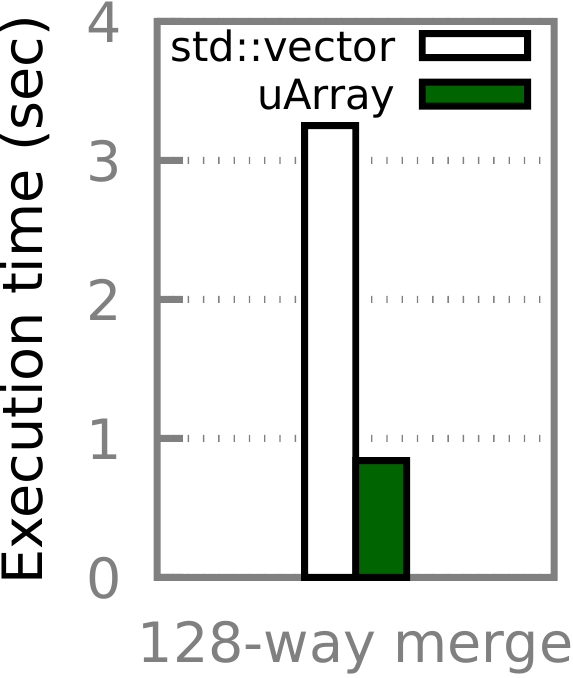}
    \vspace{-20pt}		
    \caption{
    On-demand growth of \ubuf{}s vs. std::vector
    }
    \label{fig:vector}
    \end{minipage}
    ~
    \begin{minipage}[]{0.325\textwidth}
    \centering
	\vspace{-10pt}
        \begin{subfigure}[t]{0.48\textwidth}
            \includegraphics[width=\textwidth]{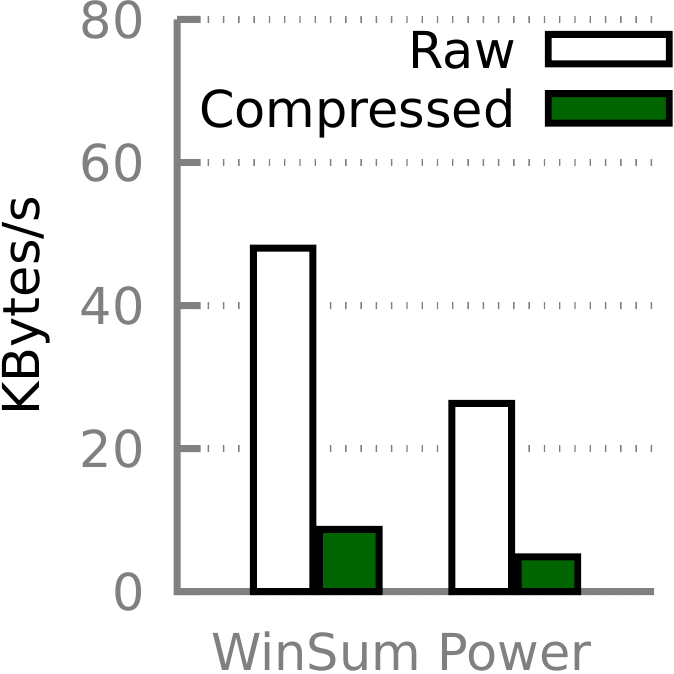}
			\caption{10K events/batch}
            \label{fig:bundle10k}
			\vspace{-5.5pt}		
        \end{subfigure}
        ~ 
        \begin{subfigure}[t]{0.47\textwidth}
    		\includegraphics[width=\textwidth]{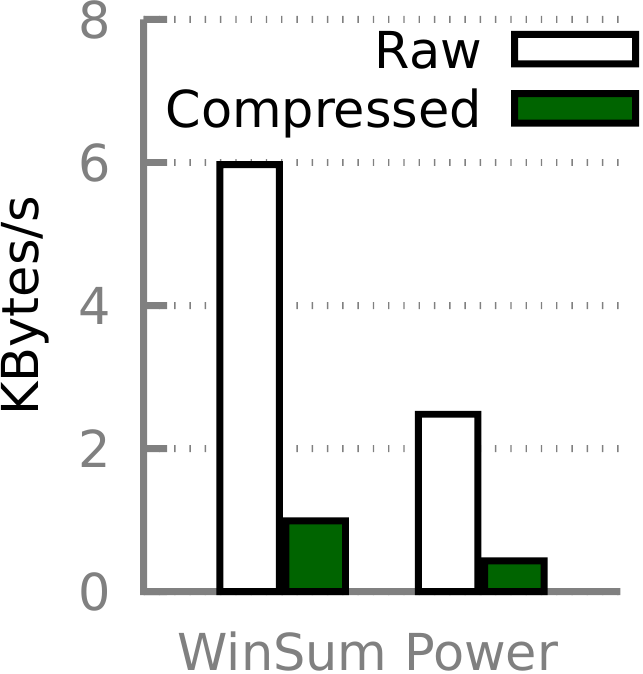}
            \caption{100K events/batch}
            \label{fig:bundle100k}
	    	\vspace{-6pt}		
        \end{subfigure}    
        \caption{Compression of audit records saves uplink bandwidth substantially. 
        }
        \label{fig:attest-exp}
    	\vspace{-10pt}		
	\end{minipage}    
   	\vspace{-18pt}		
\end{figure}

\paragraph{\ubuf{} on-demand growth}  
(\S\ref{sec:mm:abs})
We compare uArray to std::vector, a widely used C++ sequence container with on-demand growth.
We run a microbenchmark of N-way merge, an intensive procedure in \task{}s. 
It iteratively merges 128 buffers (\ubuf{}s or vectors), each containing 512 KB (128K 32-bit random integers) until obtaining a monolithic buffer;
as merge proceeds, buffers grow dynamically. 
As shown in Figure~\ref{fig:vector}, \ubuf{}s is 4$\times$ faster than std::vector,
because the allocation and paging in TEE that back \ubuf{} growth is much faster than that of a commodity OS.

\paragraph{Compression of audit records} (\S\ref{sec:attest})
The compression significantly saves the uplink bandwidth. 
We test two benchmarks (WinSum and Power) on two extremes of the spectrum of computation cost, and test two very different input batch sizes.
This is because simpler computations and smaller batch sizes generate audit records at higher rates. 
Figure~\ref{fig:attest-exp} shows that \sys{} compresses audit records by 5$\times$--6.7$\times$. 
In an offline test using gzip to compress the same records, we find our compression ratios are 1.9$\times$ higher than gzip. 
2--40 KB/sec of uplink bandwidth is saved, which is significant compared to the uploaded analytics results, which are 144 bytes/sec for WinSum and 400 bytes/sec for Power. 


\section{Related Work}
\label{sec:related}


\paragraph{Secure data analytics}
DARKLY~\cite{DARKLY} protects sensor data by isolating computations in an OS process, resulting in a large TCB.
VC3~\cite{vc3} and SecureStreams~\cite{securestreams} use SGX to protect the operators in distributed analytics. 
They lack optimizations for parallel execution in one TEE on the edge. 
To process data confidentiality, STYX~\cite{styx} computes over encrypted data, a method likely prohibitively expensive to edge platforms. 
Opaque~\cite{opaque} protects data access patterns of distributed operators, targeting a threat out of our scope.

\paragraph{TCB minimization}
Minimizing TCB is a proven approach towards a trustworthy system. 
Flicker~\cite{Flicker} directly executes security-sensitive code on baremetal hardware. 
Trustvisor~\cite{trustvisor} shrinks its TCB to a specialized hypervisor. 
Sharing a similar goal, \sys{} addresses unique challenges in supporting data-intensive computation on a minimal TCB.

\paragraph{Trusted Execution Environments}
Much work isolates security-sensitive software components. 
Terra~\cite{Terra} supports isolation with a virtual machine.
Many systems used \TZ{} and SGX~\cite{intel-sgx} for TEE.
Some systems enclose in TEE whole applications~\cite{haven,scone,graphene-sgx,trustshadow}, 
while others partition existing programs for TEE~\cite{panoply,glamdring,rubinov16icse}.
These approaches often result in larger TCBs and/or higher overhead than \sys{} and are thus less desirable for \sys{}. 
TEE also sees various novel usage, including protecting mobile app classes~\cite{tlr}, enforcing security policies~\cite{TZ-Regulating}, remote attestation of application control flows~\cite{c-flat}, and controlling data access~\cite{Graduated}.
None addresses data-intensive computation as \sys{} does. 



\noindent \textbf{Edge processing} evolves from a vision~\cite{shi16iotj,satya-edge-iot} to practice~\cite{msft-azure-edge,cisco-edge-fabric,dell-statistica}.
Most works focused on programming paradigms~\cite{spanedge}, developing and deploying application~\cite{farmbeats,wearable-assistance,airbox}, and resource management~\cite{7774351}. 
Complementary to them, \sys{} focuses on secure analytics on the edge. 

\noindent \textbf{Stream processing systems}, in response to big data challenges, evolve from single-threaded~\cite{thies2002streamit,streambase,chandrasekaran2003telegraphcq,maier2005semantics,cranor2003gigascope} to massive parallel systems~\cite{sparkstreaming,timestream,naiad,dataflowAkidau2015,heron,timestream,streamscope}.
The existing systems focus on challenges, such as fault tolerance~\cite{sparkstreaming}, fast reconfiguration~\cite{drizzle}, high parallelism~\cite{trill,streambox}, and the use of GPUs~\cite{saber}.
Few systems achieve data security and performance simultaneously as \sys{} does. 

\section{Conclusions}
This paper presents \sysfull{} (\sys{}), a secure stream analytics engine designed and optimized for a TEE on an edge platform.
\sys{} offers strong data security, verifiable results, and competitive performance. 
On an octa core ARM machine, 
\sys{} processes up to tens of millions of events per second;
its security mechanisms incur less than 25\% overhead.






\section*{Acknowledgments}

The authors thank the anonymous reviewers and our shepherd, Eyal de Lara, for their insightful comments.
For this project: the authors affiliated with Purdue ECE were supported in part by NSF Award \#1718702, \#1619075, Purdue University CPS/IoT Seed Grant Program, and a Google Faculty Award; 
the author affiliated with Northeastern University was supported in part by NSF Award \#1748334.

\bibliographystyle{bib/abbrv-minimal}
\bibliography{bib/abr-long,bib/xzl,bib/hongyu,bib/misc,bib/book,bib/security,bib/iot,bib/datacentric,bib/hp}

\end{document}